\documentclass[twocolumn]{aastex63}
\usepackage[]{units}
\usepackage[update,prepend]{epstopdf}
\usepackage{graphicx}
\usepackage{amssymb}
\usepackage{amsmath}
\usepackage{threeparttable}
\usepackage{hyperref}
\usepackage{color}

\usepackage{mathtools}

\usepackage{makecell}
\usepackage{bm}
\usepackage{CJK}

\newcommand{\kms} {\,km\,s$^{-1}$}
\newcommand{\masyr} {\,mas\,yr$^{-1}$}

\newcommand{\Msun}{\,M$_\odot$}

\newcommand{\MassSun}{$0.96\pm0.08$\Msun}
\newcommand{\MassWD}{$0.67\pm0.07$\Msun}
\newcommand{\MassGiant}{$1.31\pm0.15$\Msun}
\newcommand{\MassBPRPone}{$0.97\pm0.08$\Msun}
\newcommand{\MassJao}{$0.45\pm0.03$\Msun}
\newcommand{\MassTriple}{$1.60\pm0.15$\Msun}
\newcommand{\NGaiawb}{99,680}
\newcommand{\Foutlier}{0.20}

\mathchardef\mhyphen="2D
\shorttitle{Mass across the Hertzsprung-Russell diagram}
\shortauthors{Hwang et al.}

\begin{document}

\title{Dynamical masses across the Hertzsprung-Russell diagram}

\correspondingauthor{Hsiang-Chih Hwang}
\email{hchwang@ias.edu}
\author[0000-0003-4250-4437]{Hsiang-Chih Hwang (\begin{CJK*}{UTF8}{bsmi}黃翔致\ignorespacesafterend\end{CJK*})}
\affiliation{School of Natural Sciences, Institute for Advanced Study, Princeton, 1 Einstein Drive, NJ 08540, USA}

\author[0000-0001-5082-9536]{Yuan-Sen Ting (\begin{CJK*}{UTF8}{gbsn}丁源森\ignorespacesafterend\end{CJK*})}
\affiliation{Research School of Astronomy \& Astrophysics, Australian National University, Cotter Rd., Weston, ACT 2611, Australia}
\affiliation{Research School of Computer Science, Australian National University, Acton ACT 2601, Australia}
\affiliation{Department of Astronomy, The Ohio State University, Columbus, USA}

\author[0000-0002-9156-7461]{Sihao Cheng (\begin{CJK*}{UTF8}{gbsn}程思浩\ignorespacesafterend\end{CJK*})}
\affiliation{School of Natural Sciences, Institute for Advanced Study, Princeton, 1 Einstein Drive, NJ 08540, USA}

\author[0000-0003-2573-9832]{Joshua S. Speagle (\begin{CJK*}{UTF8}{gbsn}沈佳士\ignorespacesafterend\end{CJK*})}
\affiliation{Department of Statistical Sciences, University of Toronto, Toronto, ON M5S 3G3, Canada}
\affiliation{David A. Dunlap Department of Astronomy \& Astrophysics, University of Toronto, Toronto, ON M5S 3H4, Canada}
\affiliation{Dunlap Institute for Astronomy \& Astrophysics, University of Toronto, Toronto, ON M5S 3H4, Canada}
\affiliation{Data Sciences Institute, University of Toronto, Toronto, ON M5S 3G3, Canada}

\begin{abstract}

We infer the dynamical masses of stars across the Hertzsprung-Russell (H-R) diagram using wide binaries from the Gaia survey. Gaia's high-precision astrometry measures the wide binaries' orbital motion, which contains the mass information. Using wide binaries as the training sample, we measure the mass of stars across the two-dimensional H-R diagram using the combination of statistical inference and neural networks. Our results provide the dynamical mass measurements for main-sequence stars from 0.1 to 2\Msun, unresolved binaries and unresolved triples on the main sequence, and the mean masses of giants and white dwarfs. Two regions in the H-R diagram show interesting behaviors in mass, where one of them is pre-main-sequence stars, and the other one may be related to close compact object companions like M dwarf-white dwarf binaries. These mass measurements depend solely on Newtonian dynamics, providing independent constraints on stellar evolutionary models and the occurrence rate of compact objects. 

\end{abstract}
\keywords{stars: Hertzsprung-Russell and colour-magnitude diagrams -- stars: fundamental parameters -- binaries: general}

\section{Introduction}

Mass is arguably the most fundamental stellar parameter of stars. Masses together with metallicities of stars determine their stellar structures, surface temperatures, luminosities, stellar and chemical evolution, lifetime, and their ultimate fates. However, individual stellar masses are often derived from stellar photometry using stellar models like isochrone fitting, and the foundation of stellar models relies on some benchmark stars that have model-independent mass measurements \citep{Andersen1991,Delfosse2000,Torres2010}. Therefore, to validate and improve our understanding of stellar physics and models, it is critical to have model-independent mass measurements covering a wide range of the parameter space.

The orbital motions of binary stars directly reflect their mutual gravity and therefore their masses, providing the most reliable model-independent mass measurements (\citealt{Popper1980,Serenelli2021} and references therein). In particular, the individual masses can be derived for double-lined eclipsing binaries (e.g., \citealt{Popper1967,Burdge2022}), where the radial velocities give the mass ratio and the eclipsing light curves provide inclination constraints. For binary stars where the components are resolved and therefore not eclipsing, if their orbital periods remain sufficiently short ($\lesssim 10^2$\,yr) so that the observation can detect the orbital motion (i.e., the acceleration term), their component masses can also be derived to reasonable precision through the combination of astrometry and radial velocities \citep{Bowler2018,Brandt2019}. 

Due to their rareness, the double-lined eclipsing binary method is challenging to cover a wide range of stellar parameter space. For example, since the eclipsing probability is proportional to the radii at a fixed semi-major axis, eclipsing binaries are rare among stars with smaller radii (e.g., white dwarfs and brown dwarfs), with only a few reported cases (e.g., \citealt{Lodieu2015,Burdge2019, Martin2023}). Hence, some other techniques have been developed to measure the masses. For instance, gravitational redshifts along with stellar radius measurements can derive the masses of white dwarfs \citep{Falcon2010,Falcon2012,Joyce2018,Pasquini2019,Chandra2020b}, where the white dwarfs are not necessarily in binary systems. A single white dwarf passing a background source can lead to an astrometric microlensing event, resulting in a precise individual mass measurement \citep{Sahu2017}. In recent years, asteroseismology has become another tool to measure the mass (see review by \citealt{Chaplin2013}), in particular for solar-type stars, giants, and pulsating white dwarfs \citep{Hermes2017}, although the model dependence of the mass measurements varies for different stellar types.

Different mass measurement methods have different systematics, and cross-validations among different methods are often difficult due to the limited availability of these measurements. Furthermore, while the double-lined eclipsing binary method provides reliable masses, it is challenging to decompose the photometry of unresolved binaries into individual components, adding hurdles in using them to constrain stellar models. Therefore, the goal of this paper is to develop a homogeneous method to measure the masses of stars for the entire H-R diagram, placing the mass measurements of different stellar populations on the same scale.

In this paper, we use resolved wide binaries to measure the dynamical masses across the H-R diagram. The high-precision astrometry from the Gaia survey \citep{Gaia2016} enables the identification of millions of resolved wide binaries \citep{El-Badry2018b,Tian2020,Hartman2020, El-Badry2021}, critical to covering the entire H-R diagram. These resolved wide binaries have orbital periods longer than 
$10^3$\,yr, and Gaia captures their current snapshot orbital motions without acceleration terms. Different from the double-lined eclipsing binary method that measures the mass of an individual binary, we measure the mass as a function of the H-R diagram using ensemble of wide binaries. Even with a single snapshot of the orbit, from the ensemble of wide binaries we can marginalize over the nuisance parameters, in particular the orbital orientation and eccentricity \citep{Tokovinin2020a,Hwang2022ecc,Hwang2022twin}, with proper statistical inference. Furthermore, since wide binaries are resolved, their photometry is directly associated with the component stars and can be used for stellar model comparison without photometry decomposition.

Our measurements are purely based on the Keplerian motions and do not rely on astrophysical assumptions. Based on a similar motivation, \cite{Giovinazzi2022} have measured dynamical masses as a function of Gaia's absolute RP magnitudes using equal-mass (`twin') wide binaries. Here we use both equal-mass and non-equal-mass binaries to cover the two-dimensional H-R diagram along the BP$-$RP colors and the absolute G-band magnitudes, enabling the investigation of the global mass structures in the H-R diagram from a homogeneous determination.

The paper is structured as follows. We detail our methodology and sample selection in Sec.~\ref{sec:method-data}. In Sec.~\ref{sec:main-results}, we present the mass measurements for the simulated wide binaries and the wide binaries observed by Gaia. We discuss interesting mass features in the H-R diagram in Sec.~\ref{sec:discussion} and conclude in Sec.~\ref{sec:conclusion}. We use G and M$_{\rm G}$ interchangeably for absolute G-band magnitudes. The term `wide binary' and `wide pair' are referred to systems where two component stars are resolved by Gaia (with typical physical separations $\gtrsim 10^2$\,AU), regardless of whether they have unresolved companions or not. The term `single star' refers to the system with only one star (i.e., no close companions) in Gaia's spatial resolution unit. We use `component stars' and `individual stars' interchangeably for the unresolved system of a wide binary, and they can be either single stars or unresolved binaries. The component star with a brighter absolute G-band magnitude is referred to as the primary component, and the fainter one as the secondary. For wide binaries where both component stars are single stars, we call them `genuine wide binaries'. When specifically discussing wide binaries with an unresolved companion, we use `three-body system' or `triple systems'. In this paper, we focus on triples consisting of an unresolved binary and a resolved wide tertiary, instead of the triples where Gaia resolves all three component stars.

\section{Methodology and sample selection}
\label{sec:method-data}

\begin{table*}[]
    \centering
    \begin{tabular}{c|c|c}
         Symbol & Definition & Unit \\
         \hline
         $m$ & Mass & M$_\odot$ \\
         $e$ & Eccentricity & unitless \\
         $a$ & Semi-major axis & AU \\
         $s$ & Sky-projected separation & AU \\
         $v$ & Sky-projected orbital velocity & \kms\ \\
         $u$ & $=v \times \sqrt{s}$ & \kms\ AU$^{1/2}$ \\
         $\tilde{u}$ & $=u/\sqrt{m_{tot}}$ & \kms\ AU$^{1/2}$ \Msun$^{-1/2}$ \\
         $\mu$ & Proper motions & \masyr\ \\
         $f_e(e)$ & Eccentricity distribution & \\
         $\sigma_u$ & Measurement uncertainty of $u$ & \kms\ AU$^{1/2}$ \\
         $\sigma_{\Delta\mu}$ & Uncertainty of proper motion difference & \masyr\ \\
         
    \end{tabular}
    \caption{Table of notations.}
    \label{tab:notations}
\end{table*}

\subsection{Inferring mass from orbital velocities and separations}
\label{sec:basic-method}

For a binary with a circular orbit, its orbital velocity $v_{3D}$ (the relative velocity between two component stars) is

\begin{equation}
\label{eq:circular}
    v_{3D} = \sqrt{\frac{G m_{tot}}{a_{3D}}},
\end{equation}
where $a_{3D}$ is the semi-major axis of the circular binary, $G$ is the gravitational constant, and $m_{tot}=m_1 + m_2$ is the total mass of the binary where the component masses are $m_1$ and $m_2$. The subscript `$3D$' indicates that they are the magnitudes of full 3-dimensional vectors without any projection effects. Therefore, if $v_{3D}$ and $a_{3D}$ are known for a circular binary, the total mass of the binary can be derived by $m_{tot}=a_{3D} v^2_{3D} /G$.

In reality, full-3D magnitudes of $v_{3D}$ and $a_{3D}$ in Eq.~\ref{eq:circular} are observationally difficult to measure, and often only the components projected onto the plane of the sky can be well determined. Furthermore, binaries may have non-zero eccentricities, so the binary separation is a function of orbital phases, instead of a fixed value at the semi-major axis in the case of a circular orbit. Similar to Eq.~\ref{eq:circular}, these projected quantities and phase-dependent separations are related to the system mass. Hence, we define $u$ as:

\begin{equation}
\label{eq:u}
    u := v \times \sqrt{s} \propto \sqrt{m_{tot}},
\end{equation}
where $v$ is the two-dimensional orbital velocity, $s$ is the two-dimensional orbital separation, and both $v$ and $s$ are the components projected in plane of the sky. In this paper, the orbital separations $s$ are in units of AU, $v$ in \kms, and $u$ in units of \kms\,AU$^{1/2}$.  The definitions of variables and their units used in this paper are summarized in Table~\ref{tab:notations}.

To infer the binary mass based on the observed distribution of $u$, we build the likelihood model for $u$, which depends on the binary mass, binary orientation, and eccentricity distribution. Since our Sun is not located at a special place for these wide binaries, the orientation of binaries should be randomly distributed with respect to the Sun. As a test, we find the direction of the projected separation vectors are randomly oriented relative to the Milky Way disk for wide binaries with angular separations $>1$\,arcsec; below $\sim1$\,arcsec, wide binaries' resolvability depends on Gaia's scanning direction and therefore the orientation of detected wide binaries depends on the scanning pattern. In this paper, we use wide binaries with angular separations $>2$\,arcsec, and we adopt a random binary orientation in the inference.

The eccentricity distribution of wide binaries can be inferred from the `$v$-$r$ angle', which is the angle between the orbital velocity vector ($\vec v$) and the separation vector ($\vec s$). Using this method, \cite{Hwang2022ecc} showed that the eccentricity distribution $f_e(e)$ is close to uniform ($f_e(e)=1$) for wide binaries at $\sim10^2$\,AU, approximately thermal ($f_e(e)=2e$) at $10^3$\,AU, and then superthermal ($f_e(e)\propto e^{1.3}$) at $>10^3$\,AU. Since our sample is dominated by wide binaries at $\sim10^3$\,AU, our mass inference assumes a thermal eccentricity distribution ($f_e(e)=2e$), which is
a theoretical distribution when the ensemble of binaries uniformly populates the phase space at a fixed energy \citep{Jeans1919, Ambartsumian1937,Heggie1975,Kroupa2008,Hamilton2022,Modak2023a}. It is possible that some populations among the wide binaries do not follow the thermal eccentricity distribution (e.g., twin wide binaries, \citealt{Hwang2022twin}), but we reserve the modeling of different eccentricity distributions for future work.

\begin{figure}
    \centering
    \includegraphics[width=\linewidth]{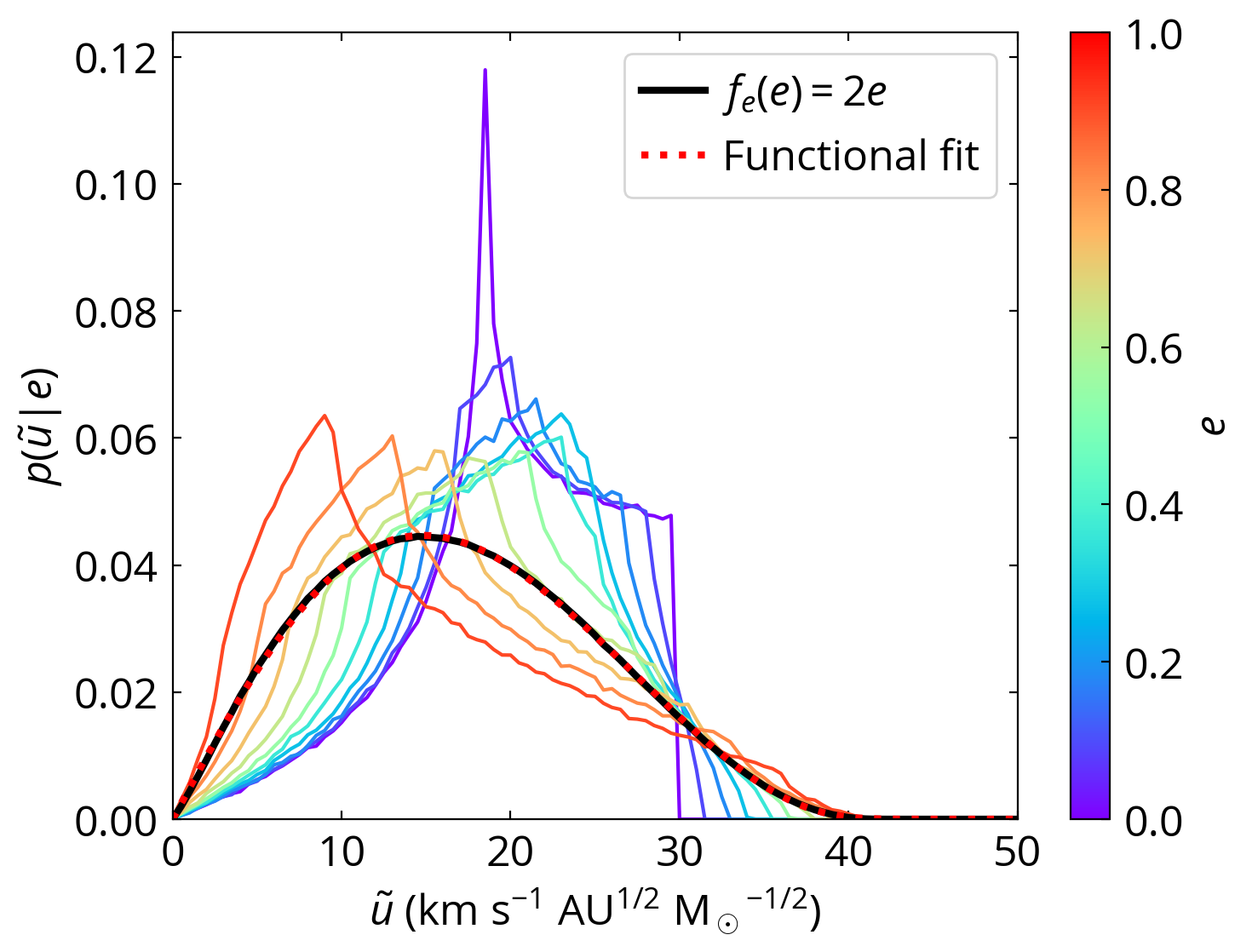}
    \caption{The simulated distributions for $p(\tilde{u}|e)$. The colored solid lines show the $p(\tilde{u}|e)$ for single-valued eccentricity $e$, where the value is represented by the color. The solid black line shows the distribution for the thermal eccentricity distribution, i.e., $p(\tilde u | f_e(e)=2e)$. The dotted red line is the functional fit using Eq.~\ref{eq:p-u-func}. }
    \label{fig:p-u}
\end{figure}

Here we define 
\begin{equation}
    \tilde{u} = u / \sqrt{m_{tot}},
\end{equation}
where the total mass $m_{tot}$ is known from simulations or measurements. Throughout the paper, the values of $\tilde u$ are in units of \kms\ AU$^{1/2}$ \Msun$^{-1/2}$. Then we derive the probability distribution functions $p(\tilde{u}|e)$ using simulations. For every eccentricity $e$ from 0 to 1 with a step of 0.01, we simulate a large number of binaries where their orbital phases are randomly sampled in time (i.e., uniformly sampled in their mean anomaly). Their binary orientations are randomly sampled so that the directions of their angular momenta are uniformly distributed on a sphere. Our code is available on GitHub\footnote{\url{https://github.com/HC-Hwang/Eccentricity-of-wide-binaries}}. $p(\tilde{u}|e)$ is numerically normalized. Fig.~\ref{fig:p-u} shows the resulting $p(\tilde{u}|e)$, where the color represents the value of $e$. For a circular orbit (purple line), its $p(\tilde{u}|e=0)$ is non-zero up to its constant circular orbital velocity $\tilde{u}=29.8$. Its strong peak at $\tilde{u}=18.4$ and the tail toward $\tilde{u}=0$ is caused by the projection effect of $s$ and $v$.

We numerically integrate $p(\tilde{u}|e)$ to derive the case for the thermal eccentricity distribution, $p(\tilde{u}|f_e=2e)$ (black line in Fig.~\ref{fig:p-u}). To accelerate the computation and to have a differentiable form for the later use of neural networks, we fit a functional form for $p(\tilde{u}|f_e=2e)$ as
\begin{equation}
\label{eq:p-u-func}
    p(\tilde{u}|f_e=2e) = A \tilde{u} \exp \left[- B\tilde{u}^2 -\exp(\frac{\tilde{u}-\tilde{u}_0}{C})\right],
\end{equation}
where $A=4.95\times10^{-3}$, $B=2.24\times10^{-3}$, $C=3.85$, and $\tilde{u}_0=36.09$. This function form is chosen simply to fit the distribution well with a minimal number of parameters. The functional fit is plotted in Fig.~\ref{fig:p-u} as the dotted red line, showing that it is well fit to the simulated black line.

For binaries with different total masses, their $p(u|e, m_{tot})$ are
\begin{equation}
    p(u|e, m_{tot}) = \frac{1}{\sqrt{m_{tot}}} p\left(\tilde u =\frac{u}{\sqrt{m_{tot}}}|e\right),
\label{eq:p-u-mtot}
\end{equation}
where the pre-factor $1/\sqrt{m_{tot}}$ ensures proper normalization. To simplify the notation, we use $p(u|m_{tot}) := p(u|m_{tot}, f_e=2e)$ and $p(\tilde u) := p(\tilde u|f_e=2e)$ for the case of the thermal eccentricity distribution.

One important property of $u$ and $\tilde u$ is that $p(u|m_{tot})$ and $p(\tilde u)$ are independent of the semi-major axis $a$. For an arbitrary binary, its $u=v\sqrt{s}$ would follow $p(u|m_{tot})$ regardless of its $a$ because the Keplerian motion is scale-free and the projection effect does not depend on $a$. Therefore, semi-major axes do not enter our likelihood later and there is no need to marginalize over the underlying semi-major axis distribution.

\subsection{From total masses to individual masses}

\label{sec:total-to-individual-mass}

In the last section, we derive the likelihood $p(u|m_{tot})$ (Eq.~\ref{eq:p-u-mtot}) so that we can infer the posterior of the total mass $p(m_{tot}|u)$ for a wide binary using the Bayes' theorem. However, how do we further constrain the masses of individual component stars? The key is that we have the photometry of individual component stars (in this paper, colors and absolute magnitudes), which provides the connection between the total mass ($m_{tot}$) and the component masses ($m_1$ and $m_2$).

For example, we can select a sample of equal-mass (twin) wide binaries where all individual components have identical colors and absolute magnitudes. Therefore, the binary mass ratios are 1, and then we can derive the individual masses from the total masses. This method is used in \cite{Giovinazzi2022}, where the authors measure the individual masses with respect to Gaia's absolute RP magnitudes.

However, twin wide binaries only constitute $<10$\%\ of all wide binaries \citep{El-Badry2019}, and there are regions in the H-R diagram (e.g., white dwarfs) where twin systems are rare. Therefore, using non-equal-mass wide binaries is necessary to cover the entire H-R diagram. Indeed, non-equal-mass wide binaries can constrain the individual masses as well. For instance, say we have three types of individual stars (star $\mathcal{A}$, $\mathcal{B}$, and $\mathcal{C}$), and there are three pairs of wide binaries, each consisting of star $\mathcal{A}$+$\mathcal{B}$, $\mathcal{B}$+$\mathcal{C}$, and $\mathcal{A}$+$\mathcal{C}$. Once we obtain the total masses of each wide binary from their orbital velocities and separations, $m_{tot,1}=m_\mathcal{A}+m_\mathcal{B}$, $m_{tot,2}=m_\mathcal{B}+m_\mathcal{C}$, and $m_{tot,3}=m_\mathcal{A}+m_\mathcal{C}$, the three individual masses can be solved by $m_\mathcal{A}=(m_{tot,1}+m_{tot,3}-m_{tot,2})/2$, etc.  This example demonstrates that non-equal-mass binaries are able to constrain the individual component masses. In reality, this procedure is more complicated because we do not deterministically measure the total mass of each wide binary. Instead, we have $p(m_{tot}|u)$ for every wide binary, and we solve the component masses through proper statistical inferences that maximize the likelihood of the observations.

Now we are ready to formulate the statistical model. For each wide binary with an index $i$, we measure its $u_i=v_i \times \sqrt{s_i}$ (Eq.~\ref{eq:u}), where $v_i$ is the projected orbital velocity and $s_i$ is the projected binary separation. Then for a sample of wide binaries, we have a set of $\{u_i\}$ from observations. Our goal is to find a function $\mathcal{M}(...)$ that maps the observables of an individual star to its individual mass. To derive the mass across the H-R diagram, we consider the BP$-$RP color and the absolute G-band magnitude measured by Gaia as the observables in this paper. Therefore, $m_{i,j} = \mathcal{M}(({\rm BP-RP})_{i,j}, {\rm G}_{i,j})$, where $m_{i,j}$ is the individual mass of the j-th ($j\in \{1, 2\}$) component star of the $i$-th wide binary, and $($BP$-$RP$)_{i,j}$ and $G_{i,j}$ are the color and the absolute G-band magnitude of the component star. Then for the i-th wide binary, its total mass is $m_{tot,i}=m_{i,1} + m_{i,2}$. The component mass we are measuring is the mass sum within Gaia's spatial resolution unit. For example, if the 1-st component star is an unresolved binary, then $m_{i,1}$ would be the total mass of the unresolved binary.

Since each wide binary is independent, we can write the likelihood as 
\begin{equation}
\label{eq:likelihood}
\log p(\{u_i\}| \mathcal{M}, \{{\rm BP-RP, G}\}_i) = \sum_i \log p(u_i|m_{tot,i}), 
\end{equation}
where $m_{tot,i}=m_{i,1}+m_{i,2}=\mathcal{M}({\rm (BP-RP)_{i,1}}, {\rm G}_{i,1})+\mathcal{M}({\rm (BP-RP)_{i,2}}, {\rm G}_{i,2})$. $p(u_i|m_{tot,i})$ can be evaluated using Eq.~\ref{eq:p-u-mtot}.

Observationally there are measurement uncertainties associated with $u$. Therefore, we marginalize over the uncertainty:
\begin{equation}
\label{eq:uncertainty}
    p_i(u_i|m_{tot,i}) = \int p_i(u'|m_{tot,i}) p(u_i|u') du',
\end{equation}
where $u$ is the observed value and $u'$ the true value, and the uncertainty distribution $p(u_i|u')$ is an assumed Gaussian distribution with a width of the associated uncertainty $\sigma_{u_i}$. Here we assume a flat prior for $p(u')$ so we can approximate $p(u_i|u')$ using $p(u'|u_i)$. We discuss $\sigma_{u_i}$ from the data perspective in Sec.~\ref{sec:data-gaia}.

Our formulation of $\mathcal{M}$ as a function of BP$-$RP and the absolute magnitude does not contain any astrophysical assumptions. For example, we do not assume any mass-temperature or mass-luminosity relation. One can include other observables of their interest, like other photometry bands and metallicity, in the parameters of $\mathcal{M}$. What we assume in the formulation is that the mass is well-behaved as a function of BP$-$RP and the absolute magnitude, and the choice of these two observables is indeed astrophysically motivated. Other than this motivation, we emphasize that our likelihood model for mass measurements is purely based on Newtonian dynamics without inputs from astrophysical models.

\subsection{Neural network model for $\mathcal{M}$}
\label{sec:NN}

There are several approaches to formulate the function $\mathcal{M}($BP$-$RP$, $G$)$. For example, one can bin the H-R diagram and assign a mass to each bin, i.e., $m[k]=\mathcal{M}({\rm BP-RP}, {\rm G})$ if (BP$-$RP, G) is in the $k$-th bin. We use this formulation along with the Markov Chain Monte Carlo (MCMC) method to derive the posterior distribution of the mass measurements in Appendix~\ref{sec:mcmc}. However, this method requires pre-defined bins on the H-R diagram, and can only work with a small number of parameters due to its expensive computation.

Our goal is to map the mass across the H-R diagram without priori knowledge like pre-defined bins. We end up using a neural network to construct the target function $\mathcal{M}({\rm BP-RP}, {\rm G})$. The advantage of using a neural network is that it does not require pre-defined bins on the H-R diagram. Instead, a neural network takes the two input parameters (BP$-$RP and G), and outputs the component mass after a series of algebraic transformations. The mass model from the neural network is trained by optimizing all the hyperparameters (weights and biases) in the network. Based on the Universal Approximation Theorem (e.g., \citealt{Cybenko1989, Funahashi1989}), a neural network can theoretically approximate any continuous functions, thus providing the most general formulation to map the observables to the mass.

We use the neural network implementation from \texttt{PyTorch} (\citealt{Pytorch2019}, v2.0.1). We use a 5-layer fully connected neural network, with 128 neurons in each layer. The Gaussian error linear units (GELU) are used for the activation function. Dropout is used during training and prediction so that each neuron has a 20\% chance of being dropped out. Dropout avoids overfitting but also provides an estimate for the model prediction uncertainty based on the dropout variational inference \citep{Kendall2017, Leung2019}. The neural network takes two inputs: the color (BP$-$RP) and the absolute magnitude (G) of a star. To ensure that the prediction mass is always positive, the neural network's output is the mass exponent $x$, and the mass is $m=\exp(x)$\Msun. During the training, there are 1024 wide binaries in each training batch, and the total number of the training sample is $\sim0.1$\,million. Different from traditional neural networks which often use the $L_2$ (mean squared errors) loss function, we use the likelihood (Eq.~\ref{eq:likelihood} with the outlier mixture model introduced in the next section) as the loss function and then adjust the model parameters using back-propagation. The Adam optimizer is used \citep{Kingma2015}. The training typically goes through the entire sample 500-1000 times (`epochs'). During the prediction, we use the trained model to generate 1000 predicted masses for a given star with dropout activated, and we use the median of the predicted masses as the reported mass and the standard deviation as the reported mass uncertainty. The training is done on the Google Colab platform with graphics processing units (GPU) to accelerate the computation. The training of 1000 epochs takes about 15 minutes using a V100 GPU.

\subsection{Wide binaries from Gaia}
\label{sec:data-gaia}

\begin{figure*}
    \centering
    \includegraphics[width=0.49\linewidth]{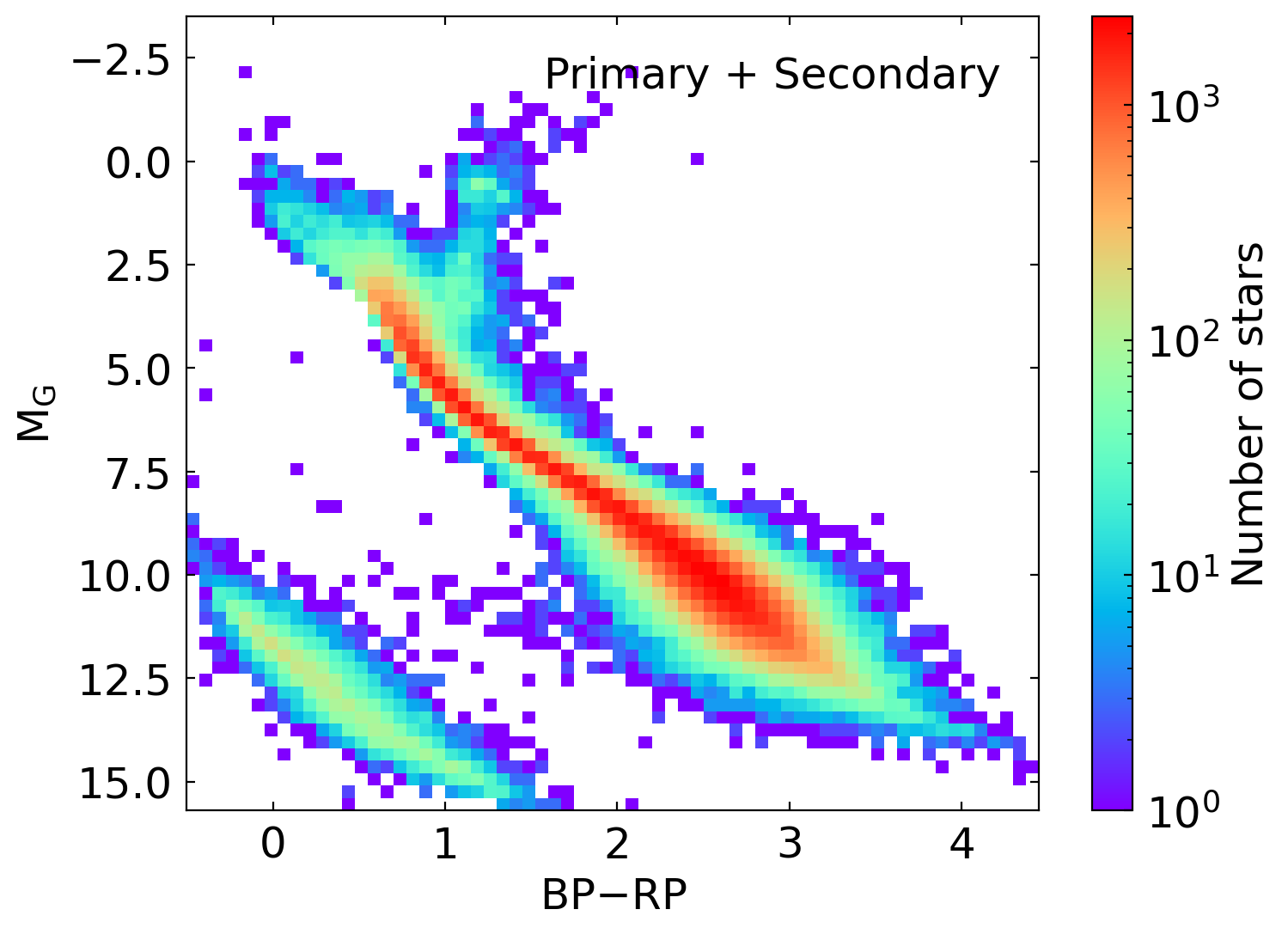}
    \includegraphics[width=0.49\linewidth]{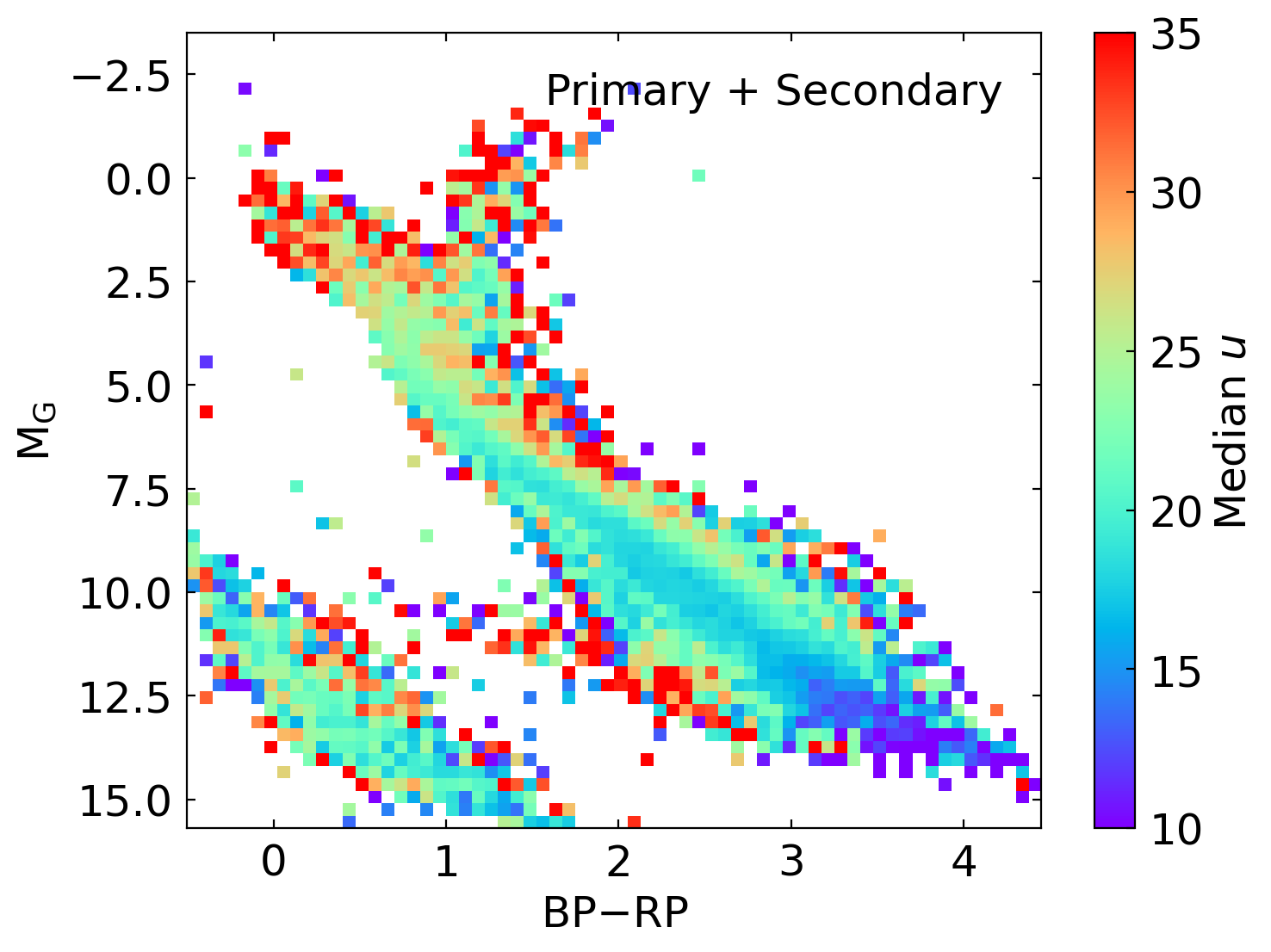}
    \caption{Left: the number of wide binaries in our sample, including both primary and secondary stars. Right: the median $u$ (in units of \kms\ AU$^{1/2}$) across the H-R diagram, where $u$ is positively correlated with the mass in each bin. Several features of $u$ (and therefore mass) are directly noticeable, including the higher masses in blue main-sequence stars, the lower masses in red main-sequence stars, and the higher-mass unresolved binaries (and triples) parallel to the main sequence.   }
    \label{fig:gaia-data}
\end{figure*}

We use the wide binary catalog from \cite{El-Badry2021}, where the resolved wide binaries were searched using the Gaia Early Data Release 3 (EDR3; \citealt{Gaia2016,Gaia2018Brown,Gaia2021Brown}) out to 1\,kpc from the Sun. These wide binaries were identified such that the component stars of a binary have parallaxes and proper motions consistent with being a gravitationally bound system. Wide binaries in clustered regions and resolved triples (where all three components are resolved by Gaia) were excluded from the catalog. For each wide binary, \cite{El-Badry2021} estimated the probability of being a chance-alignment pair ($R_{chance}$) by doing the wide binary search at random coordinates where physical wide binaries are not expected. We adopt $R_{chance}<0.01$ to avoid chance-alignment pairs, and the separations of the wide binaries of our interest are $\sim10^3$\,AU where the contamination from chance-alignment pairs is negligible. 

For each Gaia wide binary, we measure its projected physical separation $s_i$ (in AU) and projected orbital velocity $v_i$ (in \kms) to compute $u_i=v_i\times \sqrt{s_i}$ (Eq.~\ref{eq:u}). The projected separation is $s_i= 1000\theta_i/ \varpi_i$, where the angular separation $\theta_i$ is in arcsec and the parallax $\varpi_i$ is in mas. Following \cite{El-Badry2021}, we use the parallax of the primary (the brighter component) of the wide binary to avoid possible parallax systematics in fainter sources. We require parallaxes over error $>10$. The orbital velocity is $v_i=4.74 \times \Delta \mu_i / \varpi_i$, where the proper motion difference is $\Delta \mu_i=|\vec{\mu}_{i,2} - \vec{\mu}_{i,1}|$ (in \masyr) and $\vec{\mu} = (\mu_{\alpha}^*, \mu_{\delta})$. The uncertainty of proper motion difference is\begin{equation}
\sigma_{\Delta \mu} = \frac{1}{\Delta \mu}
 [(\sigma^2_{\mu^*_{\alpha,1}} + \sigma^2_{\mu^*_{\alpha,2}})\Delta\mu^2_\alpha +
 (\sigma^2_{\mu_{\delta,1}} + \sigma^2_{\mu_{\delta,2}})\Delta\mu^2_\delta
 ]^{1/2},
\end{equation}
where $\sigma_{\mu^*_{\alpha,1}}$ and $\sigma_{\mu_{\delta,1}}$ are the uncertainty of $\mu^*_{\alpha,1}$ and $\mu_{\delta,1}$, $\Delta\mu^2_\alpha=(\mu^*_{\alpha,2}-\mu^*_{\alpha,1})^2$, and $\Delta\mu^2_\delta=(\mu_{\delta,2}-\mu_{\delta,1})^2$. The subscripts `1' and `2' indicate the two component stars of the wide binary. Our error analysis could be improved by considering the correlation between the two proper motion components. However, as most wide binaries in Gaia have a small correlation, we stick to this simpler error form.

The measurement of $u$ depends on angular separations, parallaxes, and proper motion differences. Typically, angular separations can be determined better than 0.1\% precision, and parallaxes better than a few percent precision. Therefore, the uncertainty of $u$ is dominated by the proper motion difference, and we compute the uncertainty of $u$ by $\sigma_u = u \sigma_{\Delta \mu}/\Delta \mu$. Based on Gaia's current proper motion precision, we select wide binaries with $s<10^{3.5}$\,AU and parallaxes $>2.5$\,mas (distances $<400$\,pc). In particular, a circular wide binary with a total mass of $1$\Msun\ at $a=3000$\,AU has an orbital velocity of 0.55\kms, corresponding to a proper motion difference of 0.3\masyr\,at parallaxes of 2.5\,mas. For typical proper motion uncertainties of $0.1$\masyr\ in Gaia EDR3, this sample selection ensures that most of the wide binaries can have signal-to-noise ratios (SNR) of proper motion difference $\Delta \mu/\sigma_{\Delta_\mu}>3$, and therefore $u/\sigma_u>3$.

Observational data often exist with some outliers from various origins. For instance, if a star has a close stellar companion with orbital periods of several months, it may affect the astrometric measurements derived from the single-star model \citep{Belokurov2020,Penoyre2022a,Penoyre2022b}. Indeed, the majority of the resolved tertiaries of inner astrometric binaries (identified by Gaia Data Release 3; \citealt{GaiaDR3Halbwachs2022}) are missed if the wide binary search uses the single-star solutions \citep{Hwang2023}. In this paper, we only use the single-star astrometric solutions from Gaia EDR3 (which are unchanged in DR3). To ensure that the single-star astrometric measurements are reliable, we require that both component stars have renormalized unit weight errors \texttt{ruwe}$<1.4$ \citep{Lindegren2018}. Note that Gaia DR3 only processes the astrometric binary solutions for stars having \texttt{ruwe}$>1.4$ \citep{GaiaDR3Halbwachs2022}, so our \texttt{ruwe}$<1.4$ selection effectively excludes the astrometric binaries identified in Gaia DR3. However, astrometric binaries may still be present at \texttt{ruwe}$<1.4$ \citep{Penoyre2022a}; therefore our sample may still be subject to possible outliers due to unresolved companions.

Since $p(\tilde u)$ has a sharp cutoff at $\tilde u\sim 40$ (Fig.~\ref{fig:p-u}), the likelihood model from Eq.~\ref{eq:likelihood} is vulnerable to high-$u$ outliers. A high-$u$ outlier would have an expensive penalty in the likelihood, thus overestimating the mass. To appropriately incorporate the high-$u$ outliers, we adopt a mixture model \citep[e.g.,][]{Hogg2010}:
\begin{align}
\label{eq:mixture}
    & p_{mix}(u_i|m_{tot,i}) = \\ & (1-F) p(u_i|m_{tot,i}) + F p_{outlier}(\tilde u_i=u_i/\sqrt{m_{tot,i}}), \notag
\end{align}
where $p_i(u_i|m_{tot,i})$ is Eq.~\ref{eq:p-u-mtot} for well-behaved data, and $F$ is the fraction of data that is better described by the outlier model $p_{outlier}(\tilde u)$. We adopt a Gaussian (truncated at 0) for the outlier model:
\begin{equation}
\label{eq:outlier}
p_{outlier}(\tilde u) = \frac{1}{Z} \exp \left ( -\frac{(\tilde u-\tilde u_{outlier})^2}{2\sigma_{outlier}^2} \right ),
\end{equation}
where $Z$ is the normalization factor (not exactly $\sqrt{2\pi \sigma_{outlier}^2}$ due to the truncation at 0). After trial-and-error with the Gaia data, we use $\tilde u_{outlier}=30$, $\sigma_{outlier}=20$, and $F=\Foutlier$. We also use MCMC to constrain $F$ from the data, ending up with a similar value. Our final log-likelihood becomes
\begin{equation}
\label{eq:likelihood-mix}
\mathcal{L}=\sum_i \log p_{mix}(u_i|m_{tot,i}). 
\end{equation}

The BP and RP fluxes in Gaia (E)DR3 are measured without deblending treatment, so a nearby source (wide binary companion or physically unrelated source) within $\sim2$\,arcsec would affect one star's BP and RP fluxes. Also, wide binaries with angular separations $<2$\,arcsec have significantly higher \texttt{ruwe} likely due to centroiding errors \citep{El-Badry2021}. Therefore, we require wide binaries to have angular separations $>2$\,arcsec, SNR of G/BP/RP fluxes $>10$, and each component star have \texttt{bp\_rp\_color\_excess\_factor} $<1.8$ to avoid unreliable BP$-$RP due to the nearby sources \citep{Evans2018}. We limit our sample to galactic latitudes $|b|>10$\,deg to avoid the dustier region in the thin disk, and we do not apply extinction correction for photometry. The photometric uncertainties play a minor role, and thus we do not include photometric uncertainties in our statistical model.

Fig.~\ref{fig:gaia-data} (left) shows the distribution of primaries and secondaries from the resulting \NGaiawb\ pairs of Gaia wide binaries. The mean projected separation is $1.4\times10^{3}$\,AU. These wide binaries contain main-sequence (MS) stars, (sub-)giants, and white dwarfs, covering a wide range of the parameter space in the H-R diagram for mass measurements. The right panel of Fig.~\ref{fig:gaia-data} shows the median $u$ in each bin of the left panel. Since $u\propto \sqrt{m_{tot}}$ (Eq.~\ref{eq:u}), $u$ is positively correlated with the mass in each bin, assuming the wide companion mass distribution is similar across the H-R diagram (which is a reasonable assumption but not strictly correct; \citealt{El-Badry2019}).
Therefore, $u$ in the right panel represents several features of mass, including the higher masses in blue main-sequence stars, the lower masses in red main-sequence stars, and the unresolved binaries parallel to the main sequence. This plot is a convenient way to present the data, but every $u$ measurement is actually associated with two stars at different locations (unless it is a twin) in the H-R diagram. Therefore, it is necessary to use the mass inference model introduced in the earlier sections to recover the underlying masses across the H-R diagram.

\section{Mass measurements across the H-R diagram}
\label{sec:main-results}
\subsection{Tests on mock wide binary samples}
\label{sec:mock-binary}

\begin{figure*}
    \centering
    \includegraphics[width=1\linewidth]{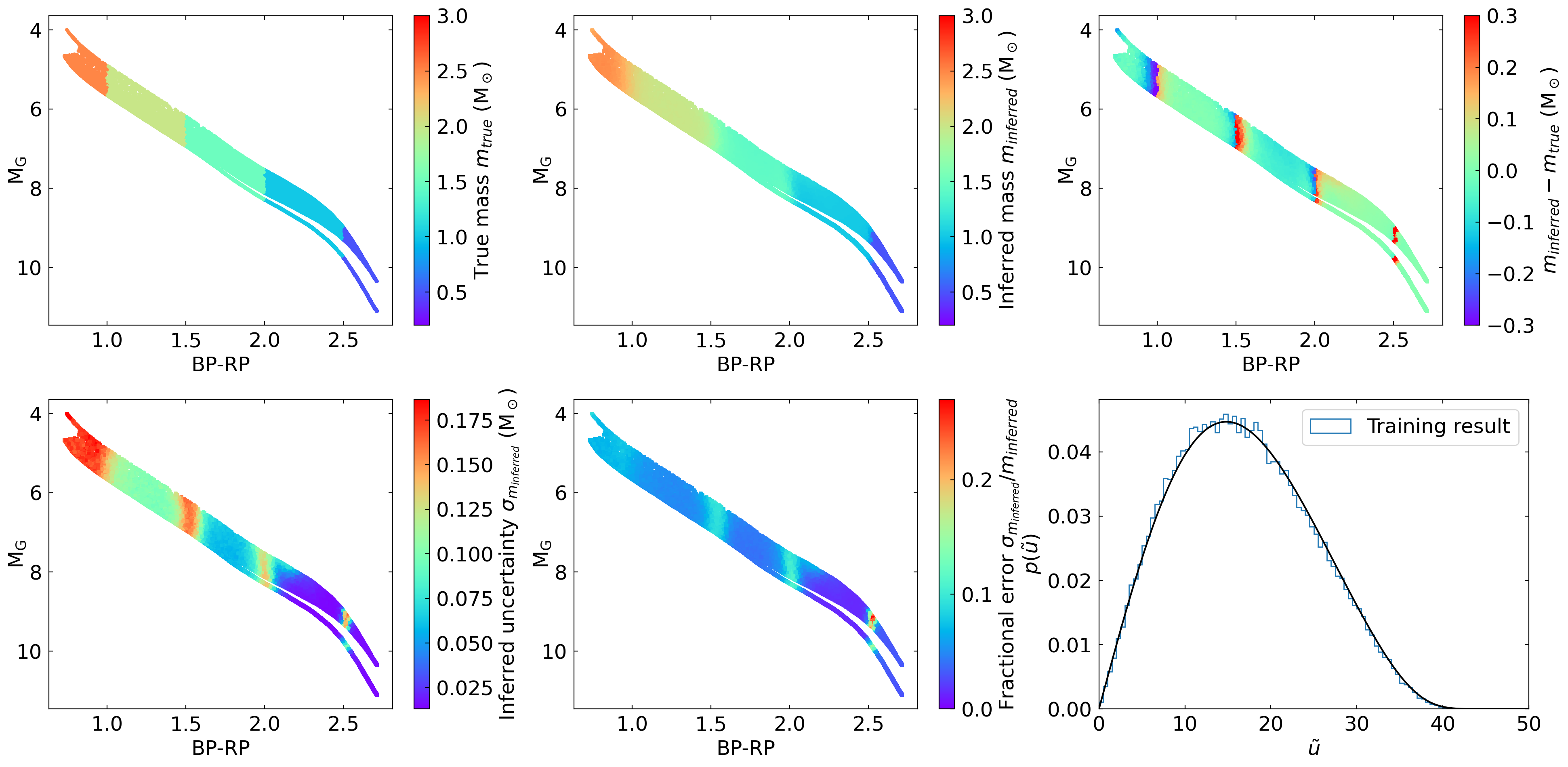}
    \caption{Mock wide binary sample with a toy model of discrete masses. First row: true mass of the sample (left), inferred mass (middle), and the difference between the inferred mass and the true mass (right). Second row: uncertainty of the inferred mass (left), the inferred fractional error (middle), and the resulting $p(\tilde u)$ distribution. }
    \label{fig:mock-binary1}
\end{figure*}

\begin{figure*}
    \centering
    \includegraphics[width=1\linewidth]{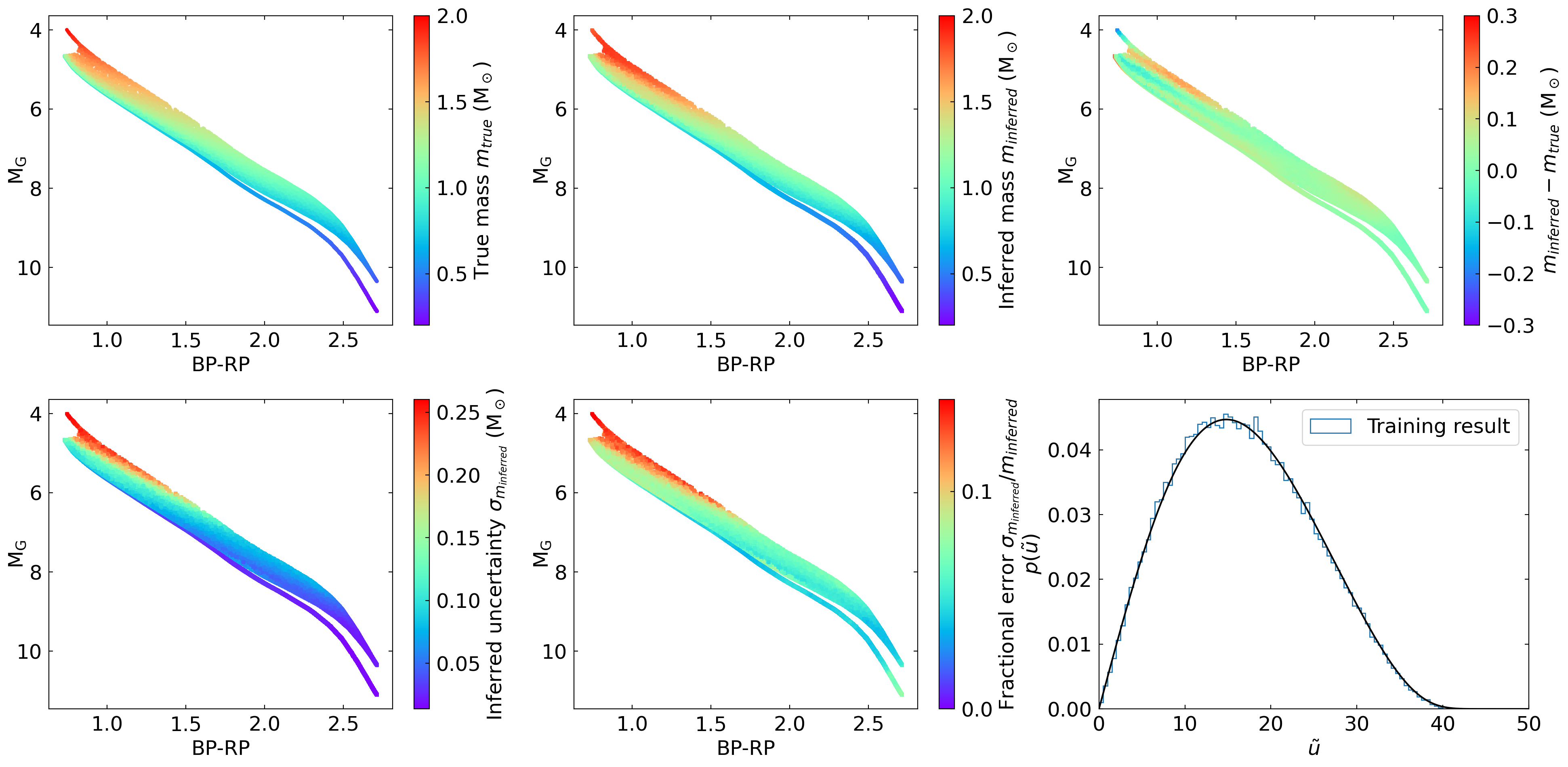}
    \caption{Mock wide binary sample with astrophysical masses. Each mock wide binary consists of two component stars, and the mass here is the mass of the component star, which can be a single star or an unresolved binary. Plot descriptions are the same as Fig.~\ref{fig:mock-binary1}. These tests show that our inference can recover the underlying mass from the orbital motions of wide binaries. }
    \label{fig:mock-binary2}
\end{figure*}

Here we demonstrate that our method can recover the underlying masses from the mock wide binary samples where their masses are known. Specifically, we simulate 100,000 mock wide pairs, similar to the number of Gaia wide binaries in our sample. Among the mock wide binaries, 50\% are genuine wide binaries without any unresolved companions, 40\% are triples consisting of one single star and one unresolved binary, and 10\% are quadruples consisting of two unresolved binaries. 20\% of the genuine wide binaries (i.e., 10\% of all pairs) are equal-mass binaries. The component masses are randomly drawn from the Salpeter initial mass function \citep{Salpeter1955} with a mass range between 0.2 and 1\Msun. We use \texttt{brutus}\footnote{\url{https://github.com/joshspeagle/brutus}} (Speagle et al. in prep) to generate their Gaia photometry (G, BP, RP) for single stars and unresolved close binaries based on the MIST isochrones \citep{Choi2016,Dotter2016}, assuming solar metallicities and stellar ages of $10^{9.5}$\,yr. For this test, we only consider main-sequence stars. Then for each wide pair, we randomly draw a $\tilde u_i$ from the distribution $p(\tilde u)$ assuming that they follow the thermal eccentricity distribution (Eq.~\ref{eq:p-u-func}), and scale it to $u_i=\tilde u_i \sqrt{m_{tot,i}}$, where $m_{tot,i}$ is the total mass (in \Msun) of the wide pair system. Here we assume no measurement uncertainties for $u_i$ and no outlier measurements. We train the neural network for 1000 epochs with a learning rate of 0.001. The tests aim to infer the mass across the H-R diagram from the information of $\{u_i\}$ and their photometry ((BP$-$RP)$_i$, G$_i$).

We start with a toy model where we manually set the masses of stars to discrete values such that their masses only depend on the BP$-$RP colors, regardless of whether they have unresolved companions or not. Specifically, the mass is $2.5-0.5k$\Msun\ for systems having $0.5+0.5 k\le$BP$-$RP$<1+0.5 k$ where $k\in [0,1,2,3,4]$. The ground-truth masses of the mock sample are shown in Fig.~\ref{fig:mock-binary1}, left panel of the first row. The double-track feature at BP$-$RP$\sim2.5$ is because the minimum main-sequence mass prohibits some parameter space of unresolved non-equal-mass binaries. Then we apply our neural-network-based method in Sec.~\ref{sec:method-data} (and set the outlier fraction $F=0$) to derive the mass. The second panel shows that the inferred masses recover the underlying masses from the mock wide binaries. The third panel is the difference between inferred and true mass. Most stars have differences close to zero, except the boundaries of discrete masses have larger deviations up to $\sim0.3$\Msun. This is a general problem that a continuous function is difficult to fit discontinuous boundaries. In the fourth panel, stars close to the boundaries have larger mass uncertainties estimated using the dropout variation, suggesting that the uncertainty estimates capture the underlying uncertainty (see also the 5th panel about the fractional errors). The last panel shows the resulting $\tilde u$ distribution using the mass measurements from the training (blue), in agreement with the likelihood model $p(\tilde u)$ from Eq.~\ref{eq:p-u-func} (black line).

Fig.~\ref{fig:mock-binary2} is the second test using the astrophysical masses from the MIST isochrones. The mass here is the mass within the spatial resolution limit, which can be the mass of a single star or the mass of an unresolved binary. The first panel shows the ground-truth masses. Astrophysically, at a fixed metallicity, the masses of stars determine the surface temperatures and therefore BP$-$RP colors, and the dependence of mass on absolute magnitudes at a fixed BP$-$RP is due to the mass ratios of unresolved binaries. The 2nd and the 3rd panels show that the inferred masses agree well with the true mass. There are larger differences in the higher-mass end because there are fewer high-mass stars due to the initial mass function. The 4th and 5th panels demonstrate that the uncertainty estimates also report reasonably higher uncertainties at the high-mass end. The last panel shows that the resulting $\tilde u$ using the inferred masses (blue) agrees well with the likelihood model from Eq.~\ref{eq:p-u-func} (black line).

The tests in Fig.~\ref{fig:mock-binary1} and \ref{fig:mock-binary2} suggest that our neural-network method can measure the dynamical masses from wide binaries across the entire H-R diagram. The tests also show that the measurement may be less reliable at some discontinuous mass transition (if any), and our method would report a larger uncertainty there. The size (100,000) of our mock binary sample is similar to our Gaia sample, demonstrating that we can obtain rich information about mass from the actual Gaia sample.

\subsection{Masses across the H-R diagram using Gaia wide binaries}
\label{sec:gaia-main-result}

\begin{figure*}
    \centering
    \includegraphics[width=\linewidth]{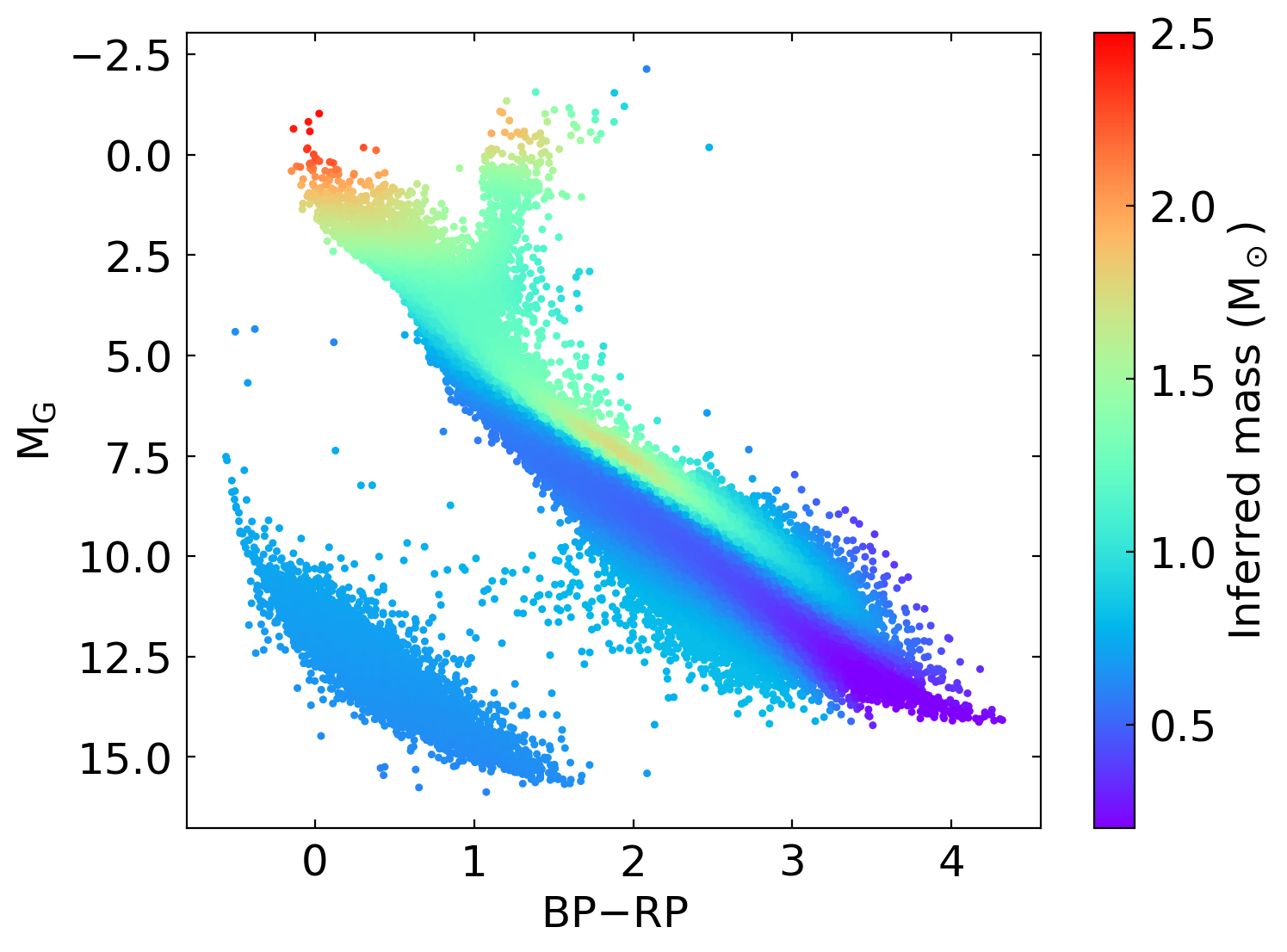}
    \caption{The inferred dynamical masses across the Hertzsprung-Russell diagram. Each point is one component star in the Gaia wide binary sample. Here we use wide binaries to represent the mass model, but we emphasize that the mass model itself is a continuous function and no data binning is involved. With our neural-network approach, we measure the masses of main-sequence stars, unresolved binaries/triples of the main sequence, (sub)-giants, and white dwarfs. An annotated version is shown in Fig.~\ref{fig:gaia-HR-mass-annotate}. }
    \label{fig:gaia-HR-mass}
\end{figure*}

\begin{figure}
    \centering
    \includegraphics[width=\linewidth]{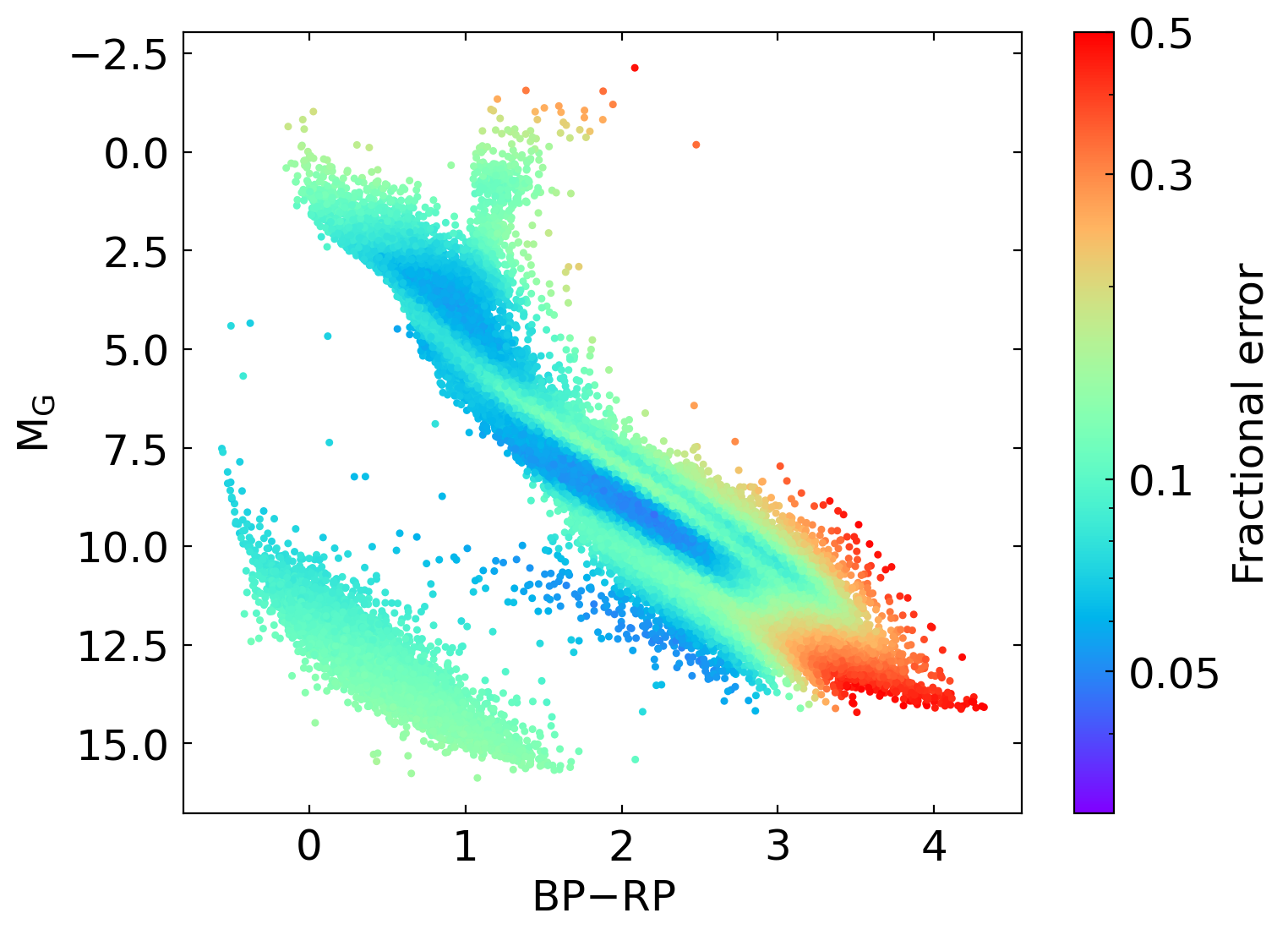}
    \caption{The estimated fractional errors of the mass measurements. The mass of most regions in the H-R diagram can be measured at $\sim10$\%\ precision. }
    \label{fig:gaia-HR-error}
\end{figure}

\begin{figure}
    \centering
    \includegraphics[width=\linewidth]{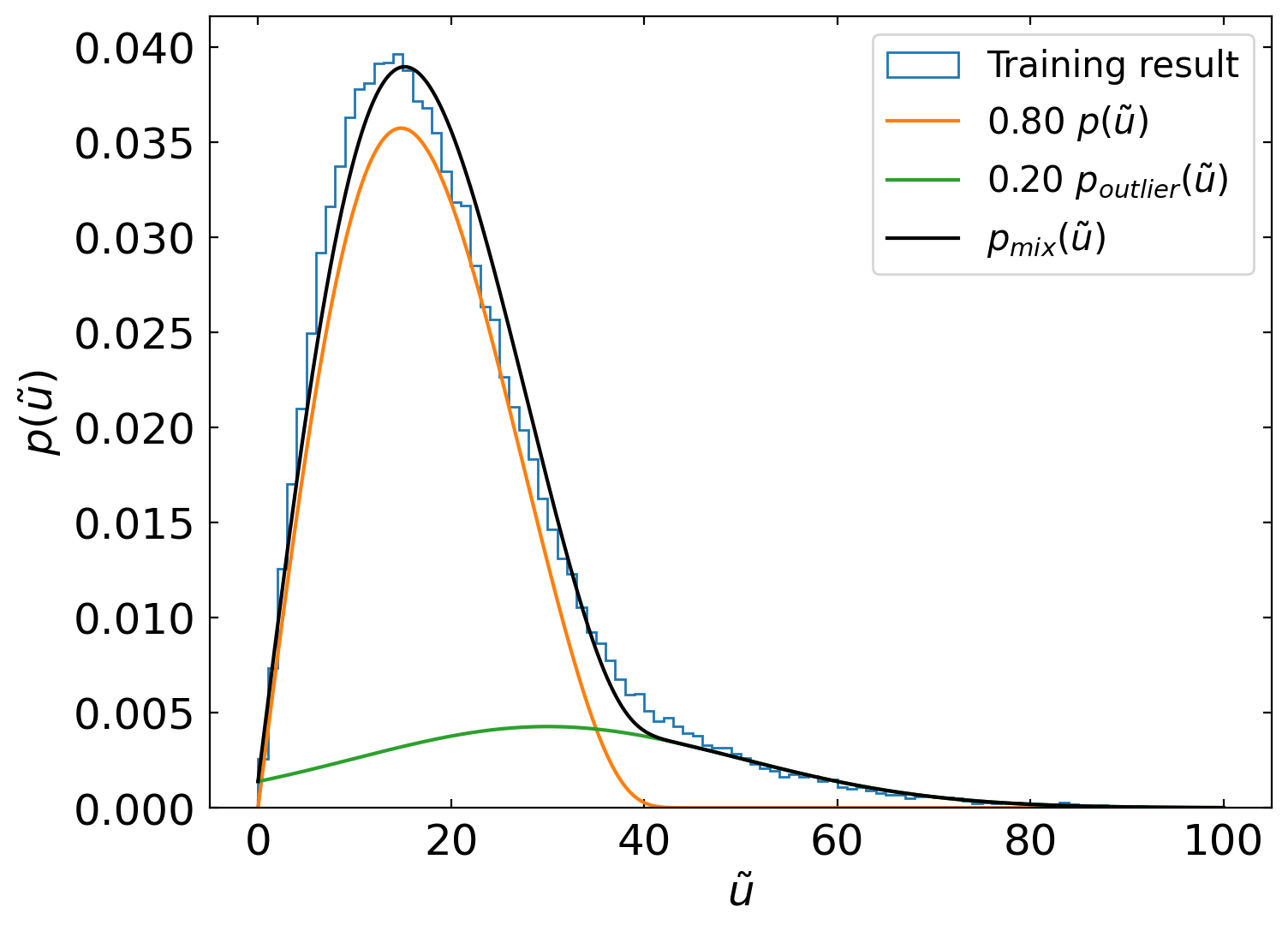}
    \caption{The resulting $p(\tilde u)$ of the wide binaries from the training (blue histogram). The orange line shows the theoretical $p(\tilde u)$ for the thermal eccentricity distribution (Eq.~\ref{eq:p-u-func}), the green line is the outlier model (Eq.~\ref{eq:outlier}), and the black line is the adopted mixture model (Eq.~\ref{eq:mixture}). The resulting $p(\tilde u)$ from the training (blue) agrees well with the mixture model (black). }
    \label{fig:gaia-HR-pu}
\end{figure}

Among the \NGaiawb\ pairs of Gaia wide binaries from Sec.~\ref{sec:data-gaia}, we randomly select 90\% of them as the training sample and 10\% as the validation sample. We then train our models using a learning rate of $1\times10^{-4}$ and an outlier fraction of $F=\Foutlier$. During the training, the average loss of the validation sample stops decreasing at epochs around 400 and becomes flat afterward. Therefore, we stop the training at 500 epochs to have a good performance in the validation sample and also avoid possible overfitting in longer epochs.

Fig.~\ref{fig:gaia-HR-mass} shows the resulting inferred dynamical masses across the Hertzsprung-Russell Diagram, where each point is a component star in the wide binary sample. Here we use these component stars to present the model, but we emphasize that the model itself is a continuous function across the H-R diagram (function $\mathcal{M}$ in Sec.~\ref{sec:total-to-individual-mass}), and there is no discrete binning involved. Only components with fractional errors $<0.5$ (99.98 per cent of the sample) are plotted. Fig.~\ref{fig:gaia-HR-error} is the inferred fractional error, showing that the masses can be measured better than 10\% in most regions of the H-R diagram. The Sun has $M_G=0.82$ and BP$-$RP$=4.67$ \citep{Casagrande2018}, where our measurement gives \MassSun. Our training result gives a mass of \MassWD\ for white dwarfs at $M_G=12$ and BP$-$RP$=0$, including all spectral types of white dwarfs. This mass is well consistent with the gravitational redshift measurements, where the mean mass of hydrogen-atmosphere (DA) white dwarfs is $0.647^{+0.013}_{-0.014}$ \citep{Falcon2010} and that of helium-atmosphere (DB) white dwarfs is $0.74^{+0.08}_{-0.09}$\Msun\ \citep{Falcon2012}.

Due to the equation of state of degenerate electrons, more massive white dwarfs are smaller in size than the less massive ones, and thus more massive white dwarfs are dimmer at fixed colors. Therefore, white dwarfs formed from single-star evolution are expected to have masses $\sim0.5$\Msun\ on the bright end and $\sim1$\Msun\ at the faint end, which is observationally confirmed through gravitational redshift measurements \citep{Chandra2020b}. In contrast, our result is consistent with a nearly constant mass around $0.70$\Msun\ for all white dwarfs. This discrepancy may be due to the fact that the number of white dwarfs is insufficient in the sample, only $\sim5$\% of the component stars. Furthermore, the mass distribution of white dwarfs strongly peaks around $0.6$\Msun \citep{Tremblay2016a}, and the higher-mass white dwarfs are rarer in wide binaries due to their significant mass loss during the late stellar evolution (Hwang et al. in preparation). Hence, there are too few high-mass white dwarfs in the wide binary sample to detect their higher masses.

Fig.~\ref{fig:gaia-HR-pu} shows the distribution of $p(\tilde u)$ of from the training result (blue). Different from the theoretical $p(\tilde u)$ (orange line), the observed $p(\tilde u)$ has a main component and an extended high-$\tilde u$ tail up to $\tilde u \sim80$. The main component agrees well with the $p(\tilde{u})$ from the thermal eccentricity distribution (orange), supporting our observationally motivated assumption that the eccentricity distribution follows the thermal eccentricity distribution \citep{Hwang2022ecc}. Overall, the behavior of the high-$\tilde u$ tail is well described by our assumed outlier model $p_{outlier}(\tilde u)$ with an outlier fraction of $F=\Foutlier$ (Sec.~\ref{sec:data-gaia}). The high-$\tilde u$ tail up to $\tilde u \sim80$ cannot be explained by eccentricities because $p(\tilde u|e)$ is strongly truncated above $\tilde u=40$ for all $e$ (Fig.~\ref{fig:p-u}) due to the small fraction of time spent at pericenter. The high-$\tilde u$ tail remains similar if we only select wide binaries with smaller errors of $u$ (e.g., $u/\sigma_u>20$). If we select wide binaries where both components have apparent G-band magnitudes brighter than 16\,mag, then the high-$\tilde u$ tail is reduced to $F\sim10$\%, suggesting that a significant fraction of the high-$\tilde u$ tail is associated with fainter sources.

The high-$\tilde u$ tail can be due to either data stochastic outliers or physical causes. For example, it can be that the astrometric quality is worse in fainter sources and is not well characterized in Gaia's reported uncertainties. Alternatively, our result assumes a one-to-one relation between the mass and the H-R diagram, and physical scenarios that do not follow this assumption may become outliers, for example the presence of compact object companions and the metallicity dependence of isochrone. Also, some outliers may be due to the astrometric noise caused by the orbital motion of the unresolved companion \citep{Penoyre2022a}. Even though the nature of the high-$\tilde u$ tail remains unclear, this paper aims to use a reasonable outlier model to take outliers into account, so that we can measure the mass through the main component reliably.

Several works have suggested that the orbital velocities of wide binaries at $\sim10^4$\,AU deviate from the expected Keplerian velocities, indicating the possible presence of non-Newtonian gravity in the low-acceleration regime (\citealt{Banik2018,Pittordis2019,Hernandez2023,Chae2023}, but see alternative explanations in \citealt{El-Badry2019b}). In Fig.~\ref{fig:gaia-HR-pu}, we show that at $\sim10^3$\,AU wide binaries, there is already a tail of higher-than-Keplerian velocities. Therefore, they may be of the same origin (e.g., worse astrometric quality in fainter sources, astrometric noise from unresolved companions, etc.), and it is just that at $\sim10^4$\,AU binary separations, the Keplerian orbits become subdominant than the noise, making the observed velocities `non-Newtonian'. Future work is necessary to compare the difference in high-velocity tail behaviors at different wide binary separations.

Based on similar merit but quite different methods, \cite{Giovinazzi2022} measure the mass with respect to absolute RP magnitudes using equal-mass wide binaries. In Appendix~\ref{sec:mcmc}, we compare our results with theirs after photometry conversion. Our trained models and the interpolated grids are available on GitHub\footnote{\url{https://github.com/HC-Hwang/HR_mass}}. Due to the stochastic nature of the training/validation sample selection and the neural network training, the mass values may not be exactly the same from different runs of training, but the differences are consistent within the inferred mass uncertainty. Therefore, the overall mass structures in the H-R diagram and our main results are unchanged in different training runs.

\begin{figure*}
    \centering
    \includegraphics[width=0.45\linewidth]{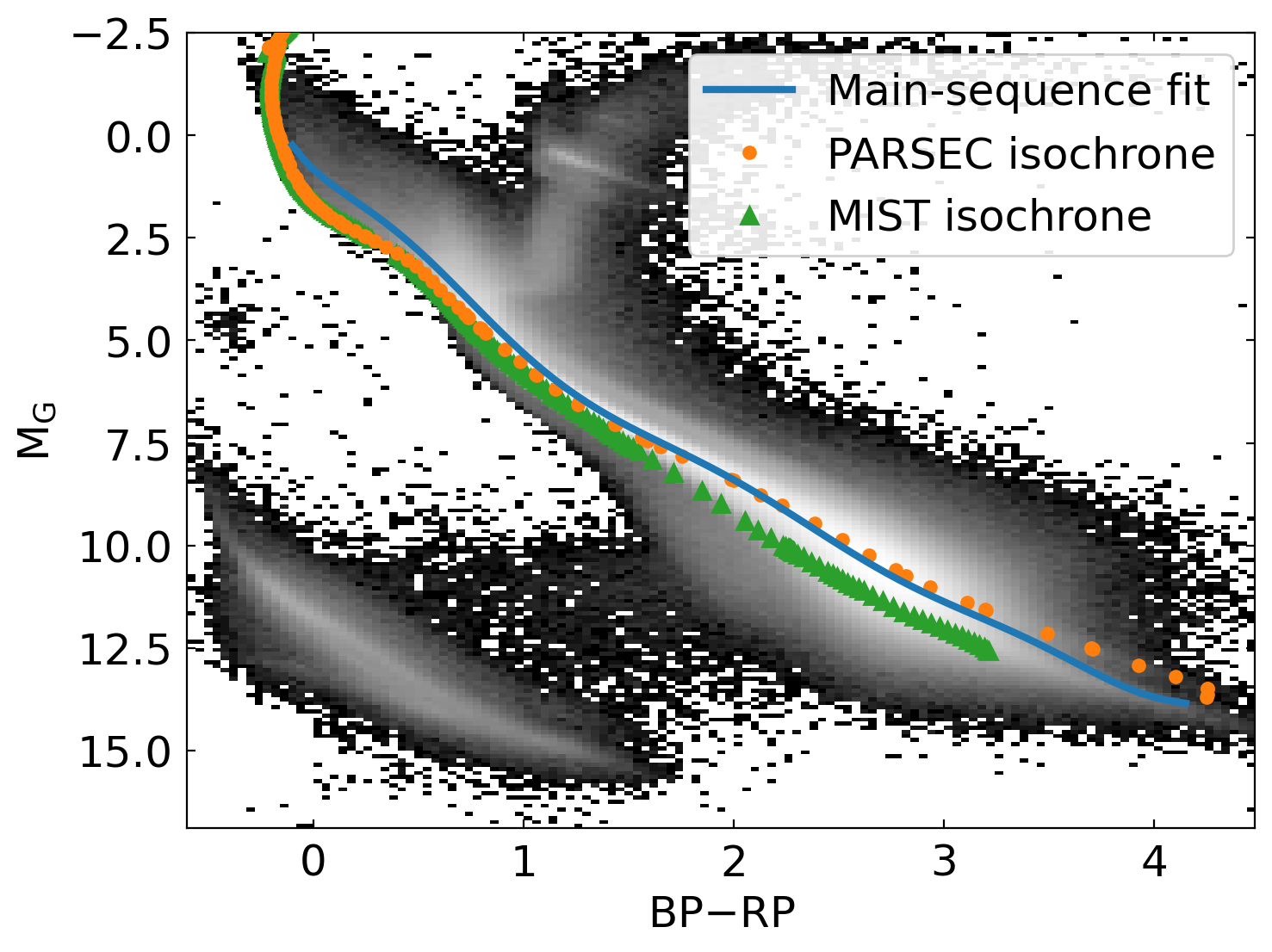}
    \includegraphics[width=0.45\linewidth]{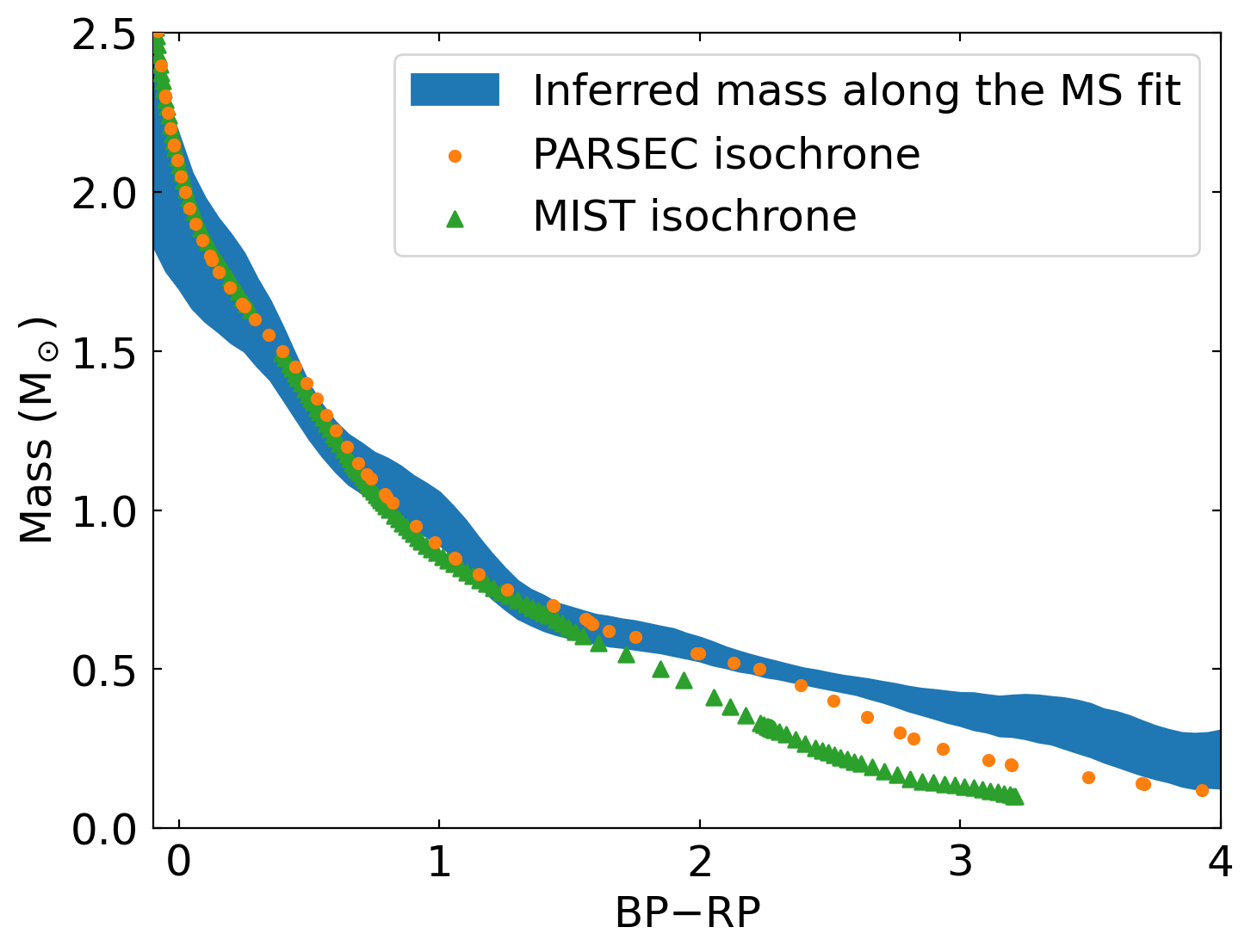}
    \caption{Comparison of masses in the single-star track. Left: the H-R diagram of the field stars. The blue line shows the main-sequence fit to the field stars. The orange and the green markers are the PARSEC and MIST isochrones respectively, with a stellar age of 0.1\,Gyr. Right: the mass as a function of BP$-$RP. The blue band is the measured masses along the main-sequence fit (blue line in the left panel). The vertical width of the line represents the 1-$\sigma$ uncertainties of the mass. The orange and green markers show the masses from the isochrone models. Our measurements agree with both isochrones at BP$-$RP$<1.5$. At BP$-$RP$>1.5$, the measured masses are more consistent with the PARSEC model.  }
    \label{fig:gaia-HR-singleMS}
\end{figure*}

\begin{figure*}
    \centering
    \includegraphics[width=0.45\linewidth]{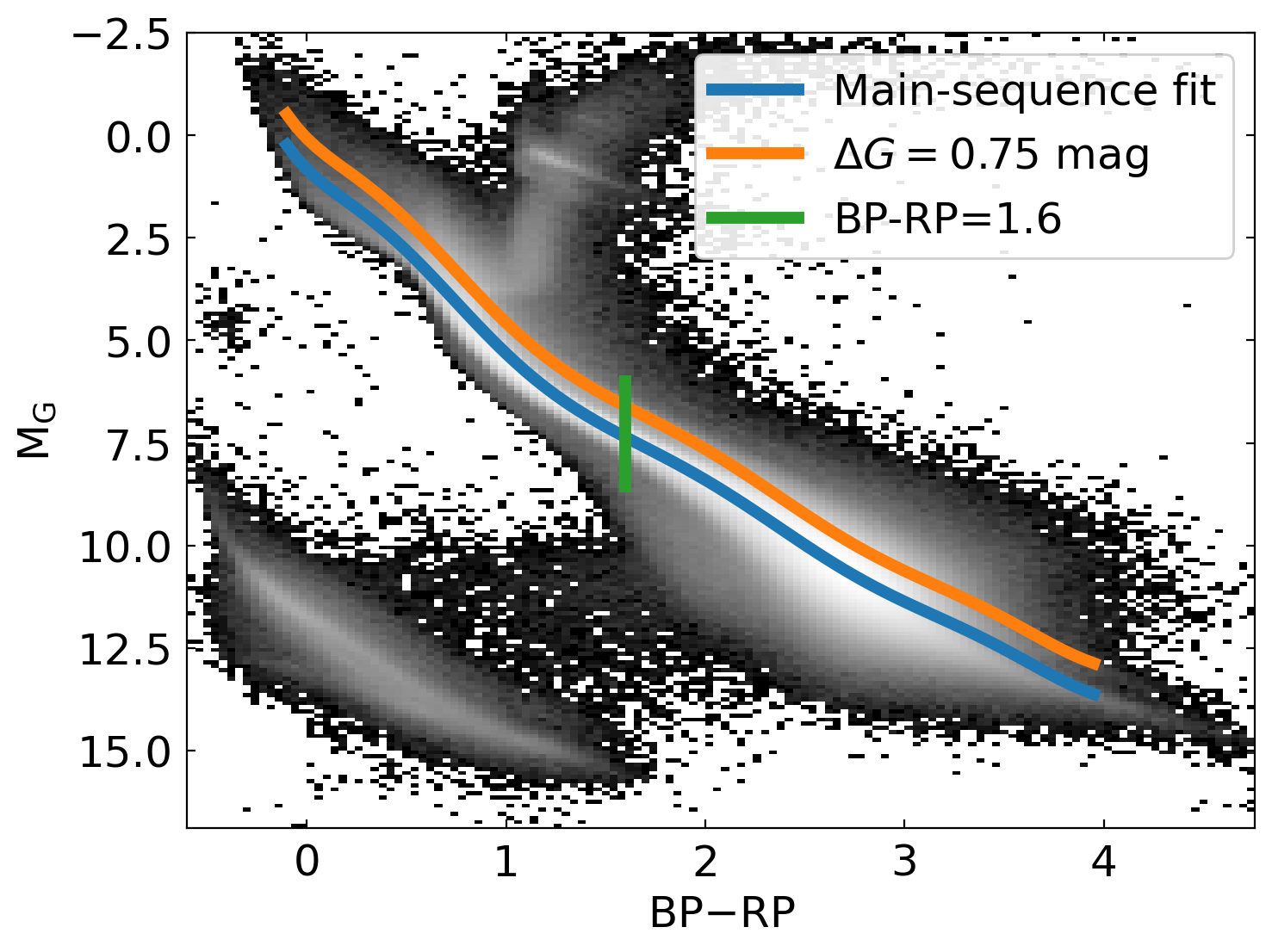}
    \includegraphics[width=0.45\linewidth]{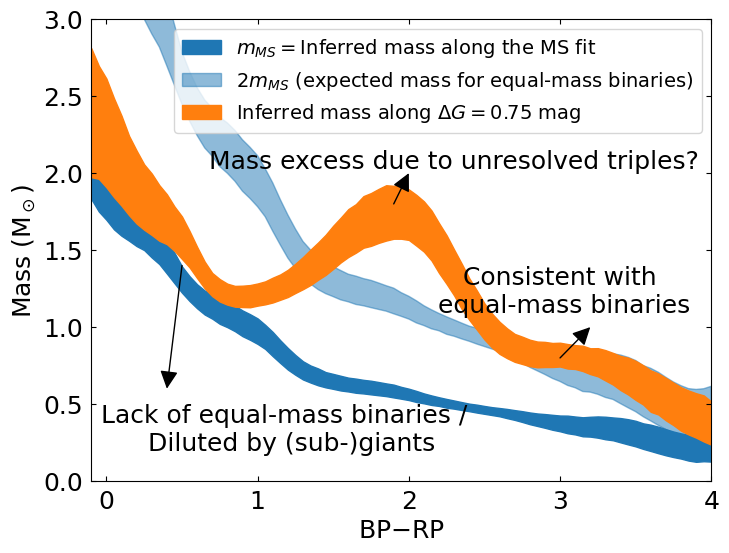}
    \caption{Comparison of masses in the unresolved binary/triple track. Left: the H-R diagram of the field stars. The blue line is the fit to the main-sequence stars (same as Fig.~\ref{fig:gaia-HR-singleMS}). The orange line is the blue line offset brighter by $\Delta G=0.75$\,mag, the expected magnitude offset for unresolved equal-mass binaries. We investigate the mass along the green vertical line in Fig.~\ref{fig:gaia-HR-vertical}. Right: the mass as a function of BP$-$RP. The dark blue line shows the mass along the single-star main-sequence track, and the light blue line shows the double mass of the dark blue line. The inferred masses along the $\Delta G=0.75$\,track (orange) show that the inferred mass is consistent with equal-mass binaries at BP$-$RP$>2.5$, and that there is a mass excess at BP$-$RP$\sim2$ that may be due to unresolved triples. }
    \label{fig:gaia-HR-twinMS}
\end{figure*}

\begin{figure}
    \centering
    \includegraphics[width=\linewidth]{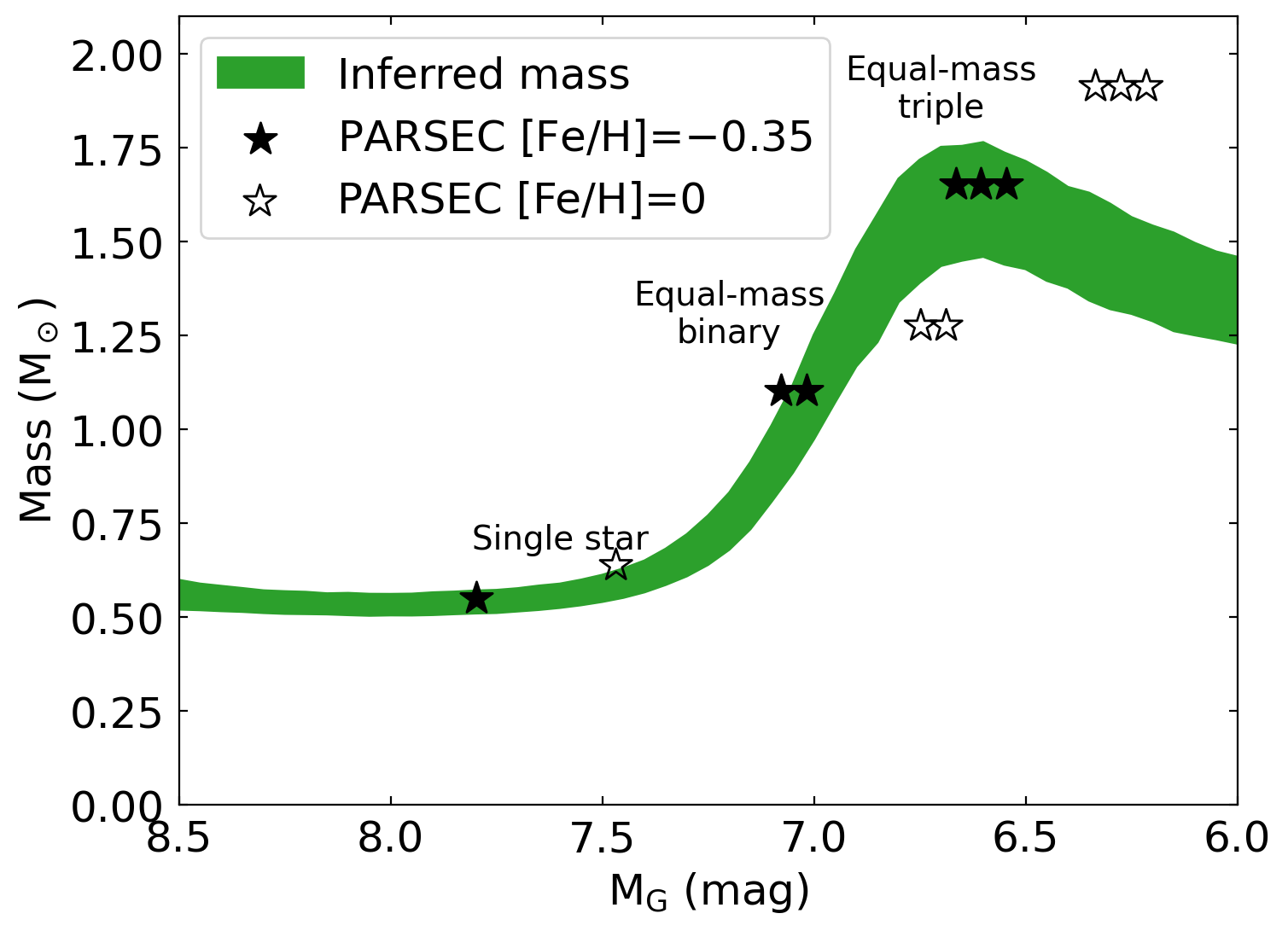}
    \caption{Mass measurements (green line) with respect to the absolute G-band magnitude (brighter to the right direction) at a fixed BP$-$RP$=1.6$ (green vertical line in Fig.~\ref{fig:gaia-HR-twinMS}, left panel). The single-star symbol represents the absolute magnitude and the mass of a single star. The double-star and triple-star symbols are for unresolved equal-mass binaries and unresolved equal-mass triples. The filled star symbols consider PARSEC models with [Fe/H]$=-0.35$, and the open star symbols for [Fe/H]$=0$. The mass peaking at \MassTriple\ along with its brighter absolute magnitude suggests a population of unresolved triples at this location of the H-R diagram. }
    \label{fig:gaia-HR-vertical}
\end{figure}

\begin{figure}
    \centering
    \includegraphics[width=\linewidth]{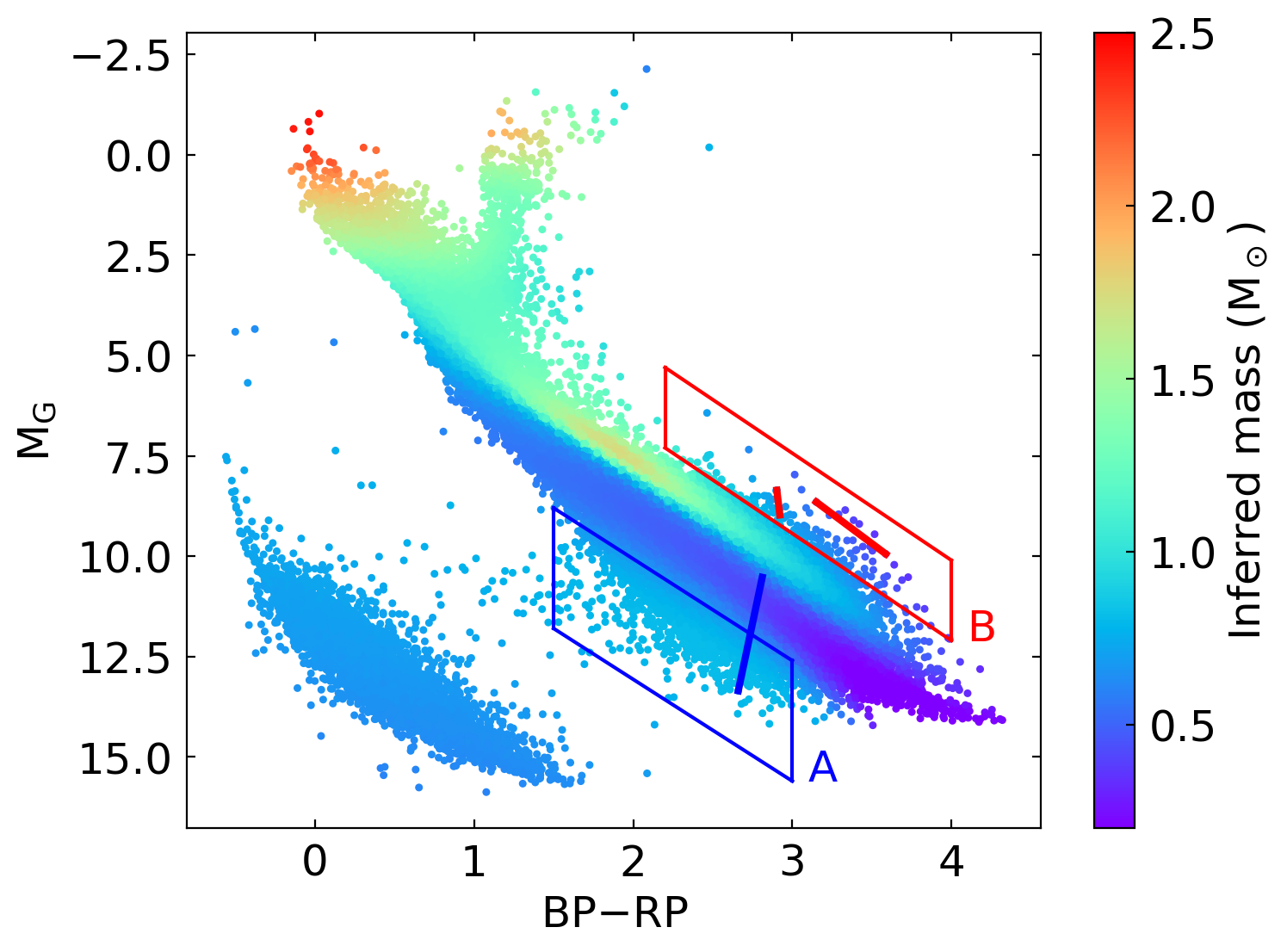}
    \caption{Two regions (A and B) showing interesting behaviors in the inferred masses. The straight lines are example wide binaries associated with these regions, where the endpoints of the line are the component stars of the wide binary. The higher mass in Region A may be associated with unresolved M dwarf-white dwarf binaries. Region B is most likely due to the pre-main-sequence wide binaries. }
    \label{fig:gaia-HR-weird}
\end{figure}

\section{Discussion}
\label{sec:discussion}

\subsection{Wide binaries as representatives of field stars}

We derive the dynamical mass across the H-R diagram from wide binaries, but is this relation applicable for general field stars? If wide binaries are a representative sampling of field stars, then the derived relation between mass and the H-R diagram applies to the field stars as well. Then the question becomes whether stars in wide binaries differ from the field stars that are not in wide binaries.

Metallicity plays an important role in determining an isochrone. We find that the wide ($10^3$-$10^4$\,AU) binary fraction peaks around the solar metallicity \citep{Hwang2021a,Hwang2022halo}. Therefore, the mean metallicity of wide binaries is similar to that of the field stars in the solar neighborhood, and they are expected to follow a similar isochrone.

Field stars and stars in wide binaries have significantly different occurrence rates of close companions.  Compared to stars without resolved wide companions, close binaries with orbital periods $<10$\,days are more common among wide pairs (i.e., wide tertiaries) \citep{Hwang2020c,Fezenko2022,Hwang2023}. However, this difference does not affect our mass measurements as long as the close binaries with wide companions follow the same mass-HR relation as the close binaries without wide companions. There is a small subset of close binaries (e.g., contact binaries) that have interaction between two component stars and thus may change their photometry \citep{Qian2003} so they do not follow the same mass-HR relation; however, the occurrence rate of contact binaries is small ($\sim0.1$\%, \citealt{Paczynski2006,Kirk2016, Hwang2020b}), leaving a negligible effect on our measurements.

Non-interacting close companions of compact objects (black holes, neutron stars, and faint white dwarfs) strongly alter the system's total mass without changing the overall photometry. If the occurrence rate of compact object close companions is higher in wide binaries (thus forming a three-body system) than in single stars (i.e., those without wide companions), then our measured mass would be higher than that of a single star due to the mass contribution from compact objects. However, we expect the occurrence rate of compact objects to be low; hence, this potential difference in compact object occurrence rate likely has a small effect on our results. Furthermore, as discussed in Sec.~\ref{sec:future} later, the mass measurement of single-like stars provides a unique constraint on the occurrence rate of compact objects.

Therefore, we consider wide binaries as reasonable representatives of field stars, and our derived mass measurements overall apply to general field stars. 
Our approach assumes an approximately one-to-one relation between mass and the H-R diagram. According to stellar physics, this assumption is only true for low-mass ($\lesssim1$\Msun) main-sequence stars with the same metallicity and for white dwarfs. Given that the solar-neighborhood is dominated by solar-metallicity stars, our mass measurements are mainly the mass relation at solar metallicities, although it is possible that some part of the H-R diagram is dominated by stars with different metallicities (e.g., the fainter part of the main sequence may have lower metallicities, \citealt{Gallart2019,Sahlholdt2019}). In the regions of sub-giants and giants in the H-R diagram, the isochrones with different ages are overlapped, and our measurements represent their masses in an average sense.

\subsection{Masses of different stellar types}

\begin{figure*}
    \centering
    \includegraphics[width=\linewidth]{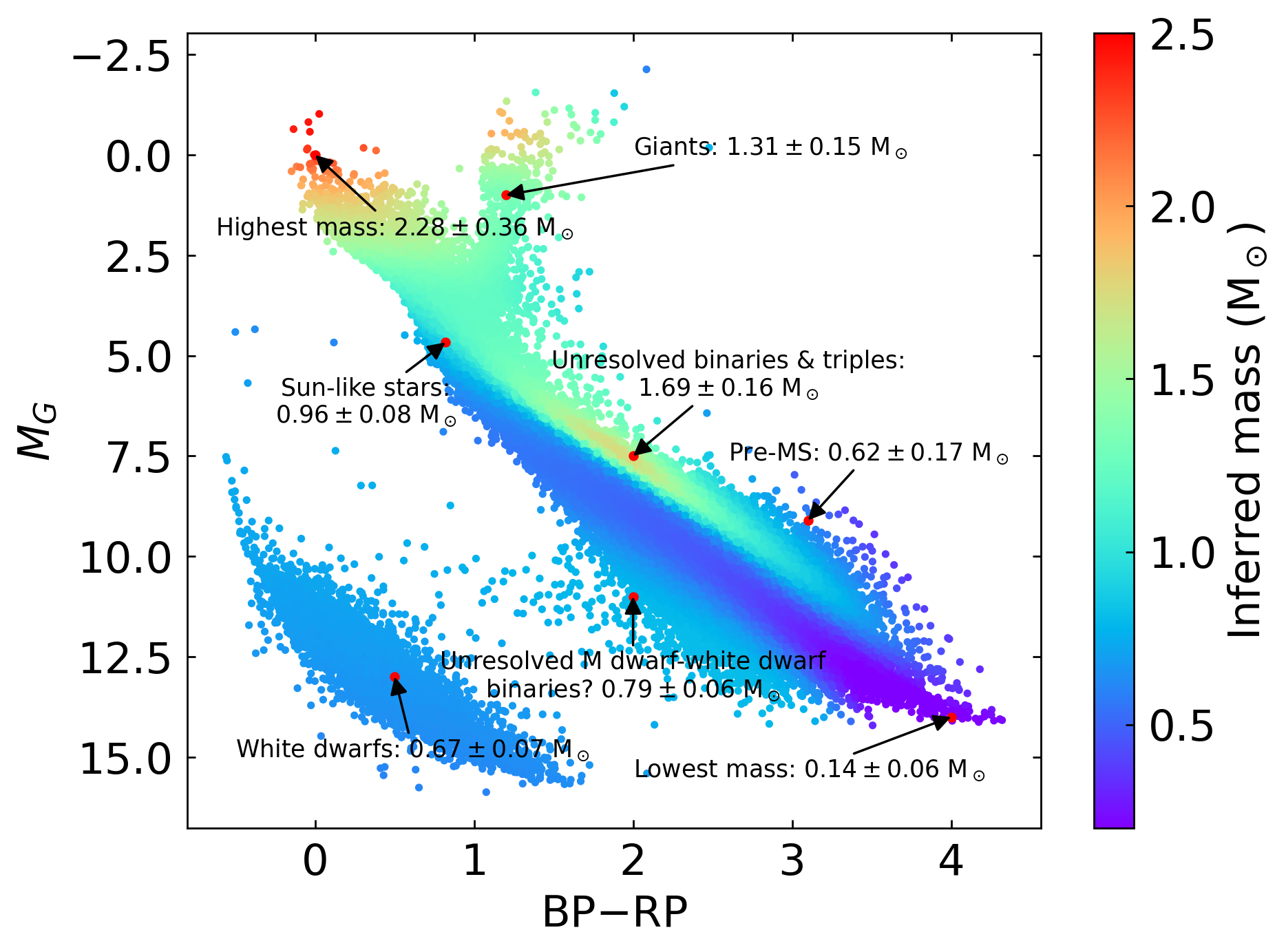}
    \caption{Same results as Fig.~\ref{fig:gaia-HR-mass}, the measured dynamical masses across the Hertzsprung-Russell diagram. Here we highlight several interesting masses in the diagram. }
    \label{fig:gaia-HR-mass-annotate}
\end{figure*}

Gaia has opened a new era of understanding the global structures of the H-R diagram. Several features are directly identified based on the density of the H-R diagram, including the gap where MS stars transition from fully to partially convective (`Jao's gap', \citealt{Jao2018}), and the anomalous accumulation of white dwarfs (`Q track') in their cooling track likely due to the crystallization at their cores \citep{Tremblay2019, Cheng2019}. In this work, we provide a different view -- the global mass structures of the H-R diagram (Fig.~\ref{fig:gaia-HR-mass} and the annotated version in Fig.~\ref{fig:gaia-HR-mass-annotate}). For example, at (BP$-$RP, G)$=(2.5,10.1)$, approximately the Jao's gap, our measured mass is \MassJao, close to (at $2.7\sigma$) the theoretical transition mass of 0.37\Msun\ for solar-metallicity stars to become fully convective \citep{Spada2017}. We note that the M-dwarf stellar models are strongly subject to the convection treatment, and thus the potential $2.7\sigma$ difference may provide an additional constraint on the stellar models. 

Fig.~\ref{fig:gaia-HR-singleMS} shows our measurements of single-MS stars compared with isochrone models. In the left panel, we fit a single-MS track (blue) which is a running median of absolute G-band magnitudes as a function of BP$-$RP (after excluding white dwarfs). Then we evaluate the mass measurements along this single-MS track in the right panel, with the width indicating $\pm 1\sigma$ uncertainties. We then compare the results with the isochrones from PARSEC (version 1.2S, \citealt{Bressan2012, Tang2014, Chen2014, Chen2015}) and from MIST (version 1.2 with rotation $v_{initial}/v_{critical}=0.4$, \citealt{Paxton2011,Paxton2013,Choi2016,Dotter2016}). We consider solar-metallicity isochrones at stellar ages of 0.1\,Gyr. Using isochrones with ages of 1 or 10 Gyr has negligible effects in the right panel, except that the high-mass ($>1$\Msun) stars are no longer on the main sequence. The right panel shows that our measurements are in good agreement with the isochrone model. At the low-mass end (BP$-$RP$>2$, or stellar masses $<0.5$\Msun), our measurements agree better with the PARSEC isochrone than MIST, providing further constraints on the stellar models of low-mass stars.

For reference, we fit the mass-BP$-$RP relation of single-MS stars (blue line in the right panel of Fig.~\ref{fig:gaia-HR-singleMS}) using 15-order polynomials. The coefficients from the highest to lowest order are: ($1.923$, $-1.918$, $6.907$, $0.839$, $-161.598$, $648.392$, $-1278.028$, $1542.543$, $-1240.595$, $692.045$, $-272.217$, $75.397$, $-14.405$, $1.809$, $-0.134$, $0.004$). The input is BP$-$RP (mag) and the output is mass (\Msun).  The fit is valid for BP$-$RP between 0 and 4.

Fig.~\ref{fig:gaia-HR-twinMS} presents our results at the unresolved binary (and triple) track. An equal-mass binary would have the same BP$-$RP colors as the single star, with a larger absolute magnitude by $\Delta G=0.75$\,mag. Therefore, we obtain the orange line in the left panel of Fig.~\ref{fig:gaia-HR-twinMS} by offsetting the single-MS line in Fig.~\ref{fig:gaia-HR-singleMS} by $\Delta G=0.75$\,mag. If unresolved equal-mass binaries dominate the sample at the $\Delta G=0.75$\,mag track (orange line), then we would expect the measured mass to be a factor of two that of the single-MS track.

In the right panel of Fig.~\ref{fig:gaia-HR-twinMS}, we compare the mass measurements on the $\Delta G=0.75$\,mag track with the single-MS track. The dark blue line is the mass of the single-star track and the light-blue line shows their double, the expected mass for the equal-mass binary. At BP$-$RP$>2.5$, the measured mass of the $\Delta G=0.75$\,mag track (orange) is consistent with two times the single MSs (light blue), suggesting that the equal-mass binaries dominate the sample. At BP$-$RP$<1$, the masses at the $\Delta G=0.75$\,mag track are consistent with those of the single-star track. This may be because the equal-mass binaries are less common in more massive stars and wide binaries \citep{Moe2017, El-Badry2019}, and also the $G=0.75$\,mag track overlap with the evolution paths of (sub-)giants, making our measurements more similar to single-star masses. 

Interestingly, at BP$-$RP$\sim2$ in the right panel of Fig.~\ref{fig:gaia-HR-twinMS}, the measured mass is larger than the expected equal-mass binaries by $0.6$\Msun\ at $3\sigma$ significance. An unresolved two-body system cannot explain the 0.6\Msun\ excess and the flux excess of $\Delta G=0.75$\,mag. Alternatively, the mass excess at this region of the H-R diagram may suggest that there is a significant fraction of unresolved triple systems.

Fig.~\ref{fig:gaia-HR-vertical} investigates more detail into the mass excess. The green line shows the measured mass versus absolute G-band magnitudes at a fixed BP$-$RP$=1.6$ (the green vertical line in Fig.~\ref{fig:gaia-HR-twinMS}, left panel). The horizontal axis of Fig.~\ref{fig:gaia-HR-vertical} is the inverted absolute G-band magnitude so that the right-hand side of the plot is brighter. The star symbols are the main-sequence masses from the PARSEC isochrones that intersect BP$-$RP$=1.6$. We consider two metallicities, a solar metallicity [Fe/H]$=0$ as the open star symbols and [Fe/H]$=-0.35$ as the filled star symbols. The single-star symbols indicate the mass and absolute magnitude for the single-star cases, the double-star symbols for unresolved equal-mass binaries, and the triple-star symbols for the unresolved equal-mass triple. At M$_G=6.6$, our inferred mass reaches the highest value of \MassTriple, exceeding the mass of equal-mass binaries for both metallicities. Since equal-mass binaries are the upper limit of mass for unresolved binaries at a fixed BP$-$RP (Fig.~\ref{fig:mock-binary2}), the highest mass of \MassTriple\ can only be explained by unresolved triples. Having a compact object companion may explain the higher mass, but cannot explain the brighter absolute magnitude. Indeed, the highest mass and its M$_{\rm G}$ agree well with the unresolved equal-mass triples at [Fe/H]$=-0.35$, where the lower metallicity may be explained by the higher close binary fraction in lower-metallicity stars \citep{Raghavan2010,Yuan2015,Badenes2018, Moe2019}. Alternatively, the highest mass of \MassTriple\ may be non-equal-mass triples with solar metallicities, which can fill the parameter space between the solar-metallicity equal-mass binary (open double-star symbol) and equal-mass triple (open triple-star symbol) in Fig.~\ref{fig:gaia-HR-vertical}.

Fig.~\ref{fig:gaia-HR-twinMS} and \ref{fig:gaia-HR-vertical} suggest that there may be a population of unresolved triples near M$_{\rm G}=6.6$ and BP$-$RP$=1.6$. Recently, hundreds of compact triple systems (where the outer tertiaries have orbits smaller than a few AU) and compact quadruples have been identified \citep{Borkovits2016,Borkovits2020a,Borkovits2020,Czavalinga2023, Rappaport2023, Kostov2023}. The formation of compact triples is particularly challenging to theory because the tertiary orbit is smaller than the radius of an initial hydrostatic stellar core \citep{Larson1969}. Therefore, our results may unveil the mysterious population of compact multiples and provide their total mass measurement.

Our mass measurements reveal two regions (A and B in Fig.~\ref{fig:gaia-HR-weird}) with interesting mass behaviors. Region A is located at the faintest part of the main sequence with BP$-$RP between 1.5 and 3. Intriguingly, stars in Region A have higher masses than the M dwarfs at similar BP$-$RP. Region A has a significantly higher $u$ in Fig.~\ref{fig:gaia-data} (right panel), in agreement with our higher inferred masses. PARSEC single-star isochrones with a low metallicity of [Fe/H]$=-1.5$ intersect Region A in the H-R diagram. However, a lower-metallicity isochrone would have a lower mass than the solar-metallicity isochrone at the same BP$-$RP, inconsistent with the observed higher mass in Region A. The mass of unresolved metal-poor binaries is still insufficient to explain the observed higher mass. Furthermore, we expect all wide binaries in our sample to be co-natal, i.e., two component stars of a wide binary have the same stellar ages and metallicities \citep{Hawkins2020, Nelson2021}; therefore, the component stars of the same wide binary would be located on the same isochrone. However, several wide binaries associated with Region A (e.g., the blue solid line in Fig.~\ref{fig:gaia-HR-weird}; its primary's Gaia DR3 source ID is 650350652407376768) are not aligned with a single metal-poor isochrone, suggesting that metallicity is not the cause for stars being in Region A. Lastly, metal-poor wide binaries are rare and are unlikely to dominate the sample in Region A \citep{Hwang2021a, Hwang2022halo}. Therefore, the origin of Region A's higher masses is not due to lower metallicities.

One possible explanation is that Region A is associated with close compact object companions. For instance, unresolved M dwarf-white dwarf binaries are located in a similar H-R diagram region \citep{Rebassa-Mansergas2021}. The unresolved white dwarf companion thus results in the higher inferred mass. This explains why some wide binaries in Region A are not aligned with any single-star isochrone, because the photometry is significantly affected by both white dwarf and M dwarf. We notice that many wide binaries associated with Region A have large $u$ (e.g., the blue solid line in Fig.~\ref{fig:gaia-HR-weird} has $u=117$) and large $\tilde u$, and Region A has a larger fraction of wide binaries associated with the high-$\tilde u$ tail (Sec.~\ref{sec:gaia-main-result} and Fig.~\ref{fig:gaia-HR-pu}). This may be because the unresolved M dwarf-white dwarf binaries in Region A have a large spread of the mass, and systems involving high-mass white dwarfs become the high-$\tilde u$ tail. The fainter absolute magnitudes of Region A may also be the reason why the high-$\tilde u$ tail is more related to the fainter sources, as discussed in Sec.~\ref{sec:gaia-main-result}. Future investigations are needed to confirm the nature of Region A and its connection with the high-$\tilde u$ tail.

In Fig.~\ref{fig:gaia-HR-weird}, Region B has brighter absolute magnitudes and lower masses than the nearby unresolved binary/triple track. Unresolved 4-body systems may explain the bright absolute magnitudes of Region B but are inconsistent with its lower inferred masses. Region B is most likely associated with pre-MS stars \citep{Zari2018, McBride2021}, when the young stars are still contracting to the main sequence. The wide binary catalog we use for mass inference has explicitly excluded clustered regions like star-forming regions \citep{El-Badry2021}, and our selection excludes systems at low Galactic latitudes ($|b|<10$\,deg). Interestingly, there are still sufficient remaining pre-MS wide binaries that dominate the sample in Region B, leaving a significant signal in our mass inference. In Fig.~\ref{fig:gaia-HR-weird}, the long solid red line shows an example of a double pre-MS wide binary (primary's Gaia DR3 ID: 49366530195371392), where both of the components have an infrared excess of W1$-$W2$=0.18$ and 0.25\,mag \citep{Wright2010,Marrese2019,Marocco2021}. This wide binary is considered a candidate member of the Taurus-Auriga star-forming complex \citep{Kraus2017}. The short solid red line is another double pre-MS wide binary (primary's Gaia DR3 ID: 6016457297110757760), where both of the components have an infrared excess of W1$-$W2$=0.39$ and 0.19\,mag. This wide binary is a candidate member of the Upper
Centaurus-Lupus region \citep{Kuruwita2018}, and its TESS light curve presents the `dipper' behaviors \citep{Capistrant2022}, which is a class of variability in young stellar objects that may be associated with circumstellar obscuration events, spots, or disk activities \citep{Cody2014}. These two examples have relatively low surrounding stellar densities, allowing them into our wide binary catalog. These young wide binaries are particularly important for testing the formation theory of wide binaries \citep{Kouwenhoven2010,Moeckel2011,Tokovinin2017,Xu2023,Rozner2023} and for understanding the multiplicity of young stars \citep{Kraus2009,Kraus2012,Kounkel2019}.

Another application of our results is the mean mass of giants. At BP$-$RP$=1.2$ and M$_G=1.0$ (Fig.~\ref{fig:gaia-HR-mass-annotate}), our measured mean mass of giants is \MassGiant, which coincides with the peak of the giant mass distribution measured from asteroseismology \citep{Yu2018, Wu2023}. The inferred mean mass of white dwarf is \MassWD, in good agreement with the literature measurements from gravitational redshifts \citep{Falcon2010,Falcon2012,Chandra2020b}. Unfortunately, we do not detect the mass gradient among white dwarfs. We discuss possible future improvements in the next section.

We summarize all the features of mass measurements in Fig.~\ref{fig:gaia-HR-mass-annotate}. With the wide binaries from the Gaia survey, we measure the dynamical masses homogeneously across the H-R diagram, covering main-sequence stars, giants, and white dwarfs.

\subsection{Caveats and future prospects}
\label{sec:future}
One caveat of our work is that we do not recover the mass gradient in the white dwarfs detected by gravitational redshifts \citep{Chandra2020b}. As discussed in Sec.~\ref{sec:gaia-main-result}, the non-detection may be due to the smaller sample sizes of high-mass white dwarfs among the wide binary sample. We have attempted to train a separate white dwarf model with no success. Future Gaia releases may be critical, improving the sample size of high-mass white dwarfs and the astrometric precision.

In recent years, there has been an exciting discovery of stellar-mass black holes in binary orbits \citep{Thompson2019, El-Badry2023BHa, El-Badry2023BHb} and as a single system from microlensing (\citealt{Sahu2022}, but see \citealt{Lam2022}). One of them is a $9.62\pm0.18$\Msun\ black hole surrounding a $0.93\pm0.05$\Msun\ Sun-like star \citep{El-Badry2023BHa}, with BP$-$RP$=1.16$, $M_G=5.35$, and metallicity [Fe/H]$=-0.2\pm0.05$. No comoving wide companion is found around the system. In principle, our mass measurements can constrain the occurrence rate of such $\sim10$\Msun\ stellar black holes among wide binaries. For example, at BP$-$RP$=1$ on the single-MS track in Fig.~\ref{fig:gaia-HR-singleMS}, our measured mass is \MassBPRPone, while the PARSEC isochrone gives a mass of 0.89\Msun, assuming a solar metallicity. If we take the isochrone mass as the ground truth of a single star's mass, we obtain an upper limit of $\sim1\%$ of the occurrence rate of 10-\Msun\ black holes among these wide binaries. We are cautious that this rough estimate assumes that the black-hole companions are not associated with the high-$\tilde u$ tail in the model, and future work is needed to investigate the nature of the high-$\tilde u$ outliers. Nevertheless, the prospect of constraining compact objects' occurrence rates is exciting, especially since this approach may be the only possible method to constrain such populations out to $\sim10^2$\,AU separations from the host stars (as long as they remain a hierarchical, stable three-body system).

Here we discuss some other caveats and prospects for this work. We adopt a thermal eccentricity distribution in the current work to perform the inference. However, it is known that the eccentricity distribution is a function of binary separations \citep{Tokovinin2020a,Hwang2022ecc}, and also that some wide-binary populations like wide twins have distinct eccentricity distributions \citep{Hwang2022twin}. Although Fig.~\ref{fig:gaia-HR-pu} suggests that the final result is consistent with the thermal eccentricity distribution as the entire sample, it is more reliable to treat eccentricity distributions as free parameters in future work. Further investigation of the nature of the high-$\tilde u$ tail and the underlying mass distribution are critical to robustly constraining compact object occurrence rates and to testing the Newtonian gravity at larger binary separations. In this work, we only consider two parameters (BP$-$RP and M$_{\rm G}$), and it would be valuable to include other astrophysical parameters like metallicities (from other spectroscopic surveys or Gaia XP spectra) to map the metallicity-dependent relation.

\section{Conclusions}
\label{sec:conclusion}
In this paper, we use wide binary stars from the Gaia survey to measure dynamical masses across the Hertzsprung-Russell diagram. Specifically, wide binaries' orbital velocities and separations measured by Gaia provide critical information for their masses (Sec.~\ref{sec:basic-method} and \ref{sec:total-to-individual-mass}), and we model the relation between mass and the H-R diagram using statistical inference and a neural network (Sec.~\ref{sec:NN}). Our neural network-based method enables us to model the mass along two dimensions (color and absolute magnitude), which is critical to cover the entire H-R diagram. Our mass measurements are purely based on Keplerian law, thus providing independent constraints for stellar evolution models. Using a mock wide binary sample, we show that our method can recover the underlying masses in the H-R diagram (Fig.~\ref{fig:mock-binary1}, \ref{fig:mock-binary2}).

The resulting mass measurements provide a global view of mass structures across the H-R diagram (Fig.~\ref{fig:gaia-HR-mass}, \ref{fig:gaia-HR-mass-annotate}). Our findings are summarized as follows:
\begin{enumerate}
    \item The inferred mass of the single main-sequence stars spans from 0.1 to 2\Msun\ (Fig.~\ref{fig:gaia-HR-singleMS}). Our mass measurements overall agree with the PARSEC and MIST isochrone models at masses $>0.6$\Msun. At lower masses $<0.6$\Msun, our measurements are more consistent with the PARSEC model.
    \item Our results show an unresolved binary/triple track parallel to the main-sequence single stars (Fig.~\ref{fig:gaia-HR-twinMS}). An additional mass excess at BP$-$RP between 1.5 and 2 suggests a population of unresolved main-sequence triples (Fig.~\ref{fig:gaia-HR-vertical}).
    \item Two regions in the H-R diagram show interesting mass behaviors (Fig.~\ref{fig:gaia-HR-weird}). Region A is located at the faint end of the main sequence at BP$-$RP between 2 and 3, with an unexpectedly higher mass of $\sim0.8$\Msun. The higher mass may be due to unresolved M dwarf-white dwarf binaries. Region B is located brighter than the nearby unresolved binary/triple track, with a smaller mass of $\sim0.6$\Msun\ than the unresolved binaries/triples. Region B is mostly likely the pre-MS stars. We have identified two example wide binaries where both component stars are in Region B, have infrared excess, and are candidate members of nearby star-forming regions.
    \item The measured mean mass of giants and white dwarfs are \MassGiant\ and \MassWD, respectively. These measurements are consistent with the literature values. The non-detection of the mass gradient among white dwarfs may be due to the lack of high-mass white dwarfs in the wide binary sample.
    \item We discuss some future improvements, including modeling the eccentricity distribution during the mass inference and a better understanding of the high-$\tilde u$ outliers. These improvements are critical to constraining the occurrence rate of compact object companions and to testing the non-Newtonian gravity theory using wide binaries.
\end{enumerate}

\section*{ACKNOWLEDGEMENTS}

The authors appreciate discussions with Josh Winn, Jo Bovy, Chris Hamilton, and Nadia Zakamska. Y.S.T. acknowledges financial support from the Australian Research Council through DECRA Fellowship DE220101520. HCH thanks the constant support and encouragement for this work from Ting-Wei Young, Nae-Chyun Chen, and Ting-Wei Liao.

{\it Facilities:} Gaia.

{\it Software:} \texttt{IPython} \citep{ipython2007}, \texttt{jupyter} \citep{jupyter2016}, \texttt{Astropy} \citep{Astropy2013,Astropy2018}, \texttt{numpy} \citep{numpy2020}, \texttt{scipy} \citep{scipy2020}, \texttt{matplotlib} \citep{matplotlib2007}, \texttt{PyTorch} \citep{Pytorch2019}.

\section*{Data Availability}

The data underlying this article are available online. The datasets were derived from sources in the public domain: Gaia Data Archive \url{https://gea.esac.esa.int/archive/}

\appendix
\restartappendixnumbering
\setcounter{figure}{0}

\section{Markov-chain Monte-Carlo with a discrete-mass model}
\label{sec:mcmc}

\begin{figure*}
    \centering
    \includegraphics[width=3in]{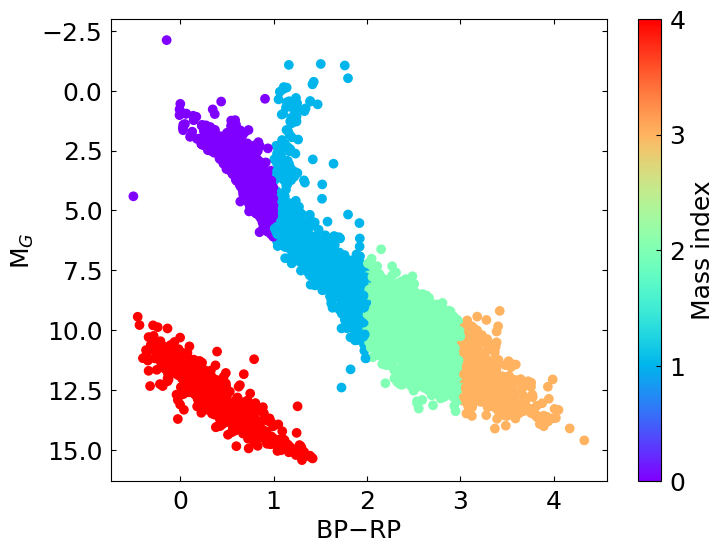}
    \caption{The H-R diagram where the color code represents their mass indices $k$ used in the MCMC method. }
    \label{fig:mass-idx}
\end{figure*}

Here we demonstrate how to use MCMC to measure the dynamical masses from wide binaries and discuss its limitations. We consider a model with 5 mass parameters $m[k]$ and assign every star an integer mass index $k\in [0,4]$ based on their locations in the H-R diagram. The color code in Fig.~\ref{fig:mass-idx} represents the assigned mass indices, where $m[4]$ is for all white dwarfs and $m[0-3]$ are for main-sequence stars with different ranges of BP$-$RP colors.
Then the total mass of wide binary $i$ is $m_{tot, i} =m[k_{i,1}]+m[k_{i,2}]$, where $k_{i,1}$ and $k_{i,2}$ are the mass indices of the component star 1 and 2. The same Gaia sample selection is used and described in Sec.~\ref{sec:data-gaia}. Among them, we randomly select 10,000 wide binary pairs to reduce the computation time. 


We use the MCMC package \texttt{emcee} \citep{Foreman-Mackey2013} to sample the posteriors of $m[k]$. The likelihood is Eq.~\ref{eq:likelihood} and Eq.~\ref{eq:outlier}, and we adopt flat priors for the mass parameters. The parameters in the outlier model is $F=0.20$, $u_{outlier}=30$, $\sigma_{outlier}=20$. We marginalize the uncertainties of $u$ using Eq~\ref{eq:uncertainty}, where we approximate the integration by a discrete summation from $u'=0$ to $u'=100$ with a step of $\Delta u'=1$. Binaries with uncertainties $\sigma_u<\Delta u'$ have a narrow $p(u_i|u')$ that would be undersampled, so we manually set their $\sigma_u$ to $\Delta u'=1$, which does not affect the main result because $p(\tilde u)$ is a smooth distribution on the scale of $\Delta u'$.

\begin{figure*}
    \centering
    \includegraphics[width=5in]{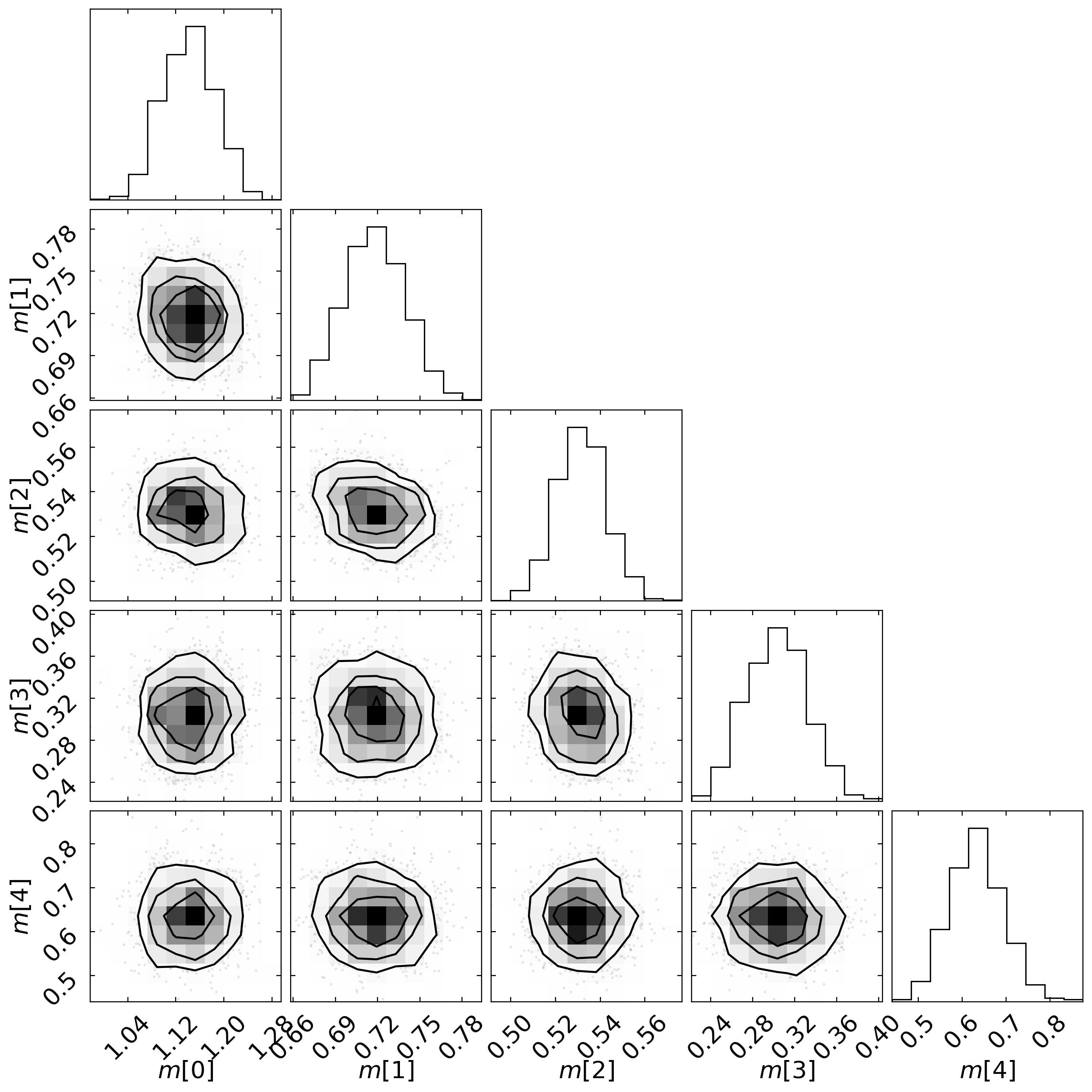}
    \caption{The posterior distribution of the mass parameters in the discrete mass model.}
    \label{fig:mcmc-corner}
\end{figure*}

    

\begin{figure*}
    \centering
    \includegraphics[width=3in]{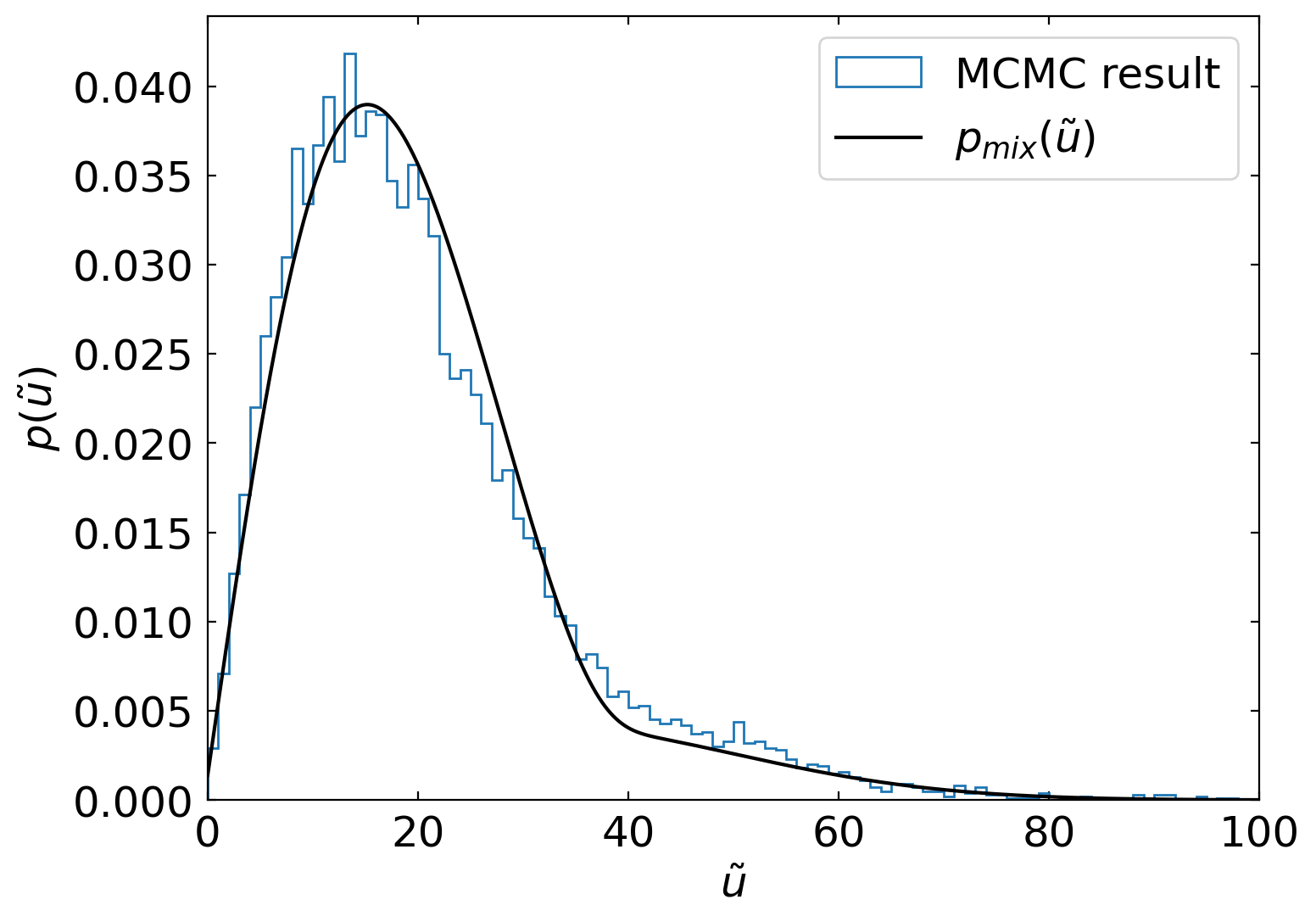}
    \includegraphics[width=3in]{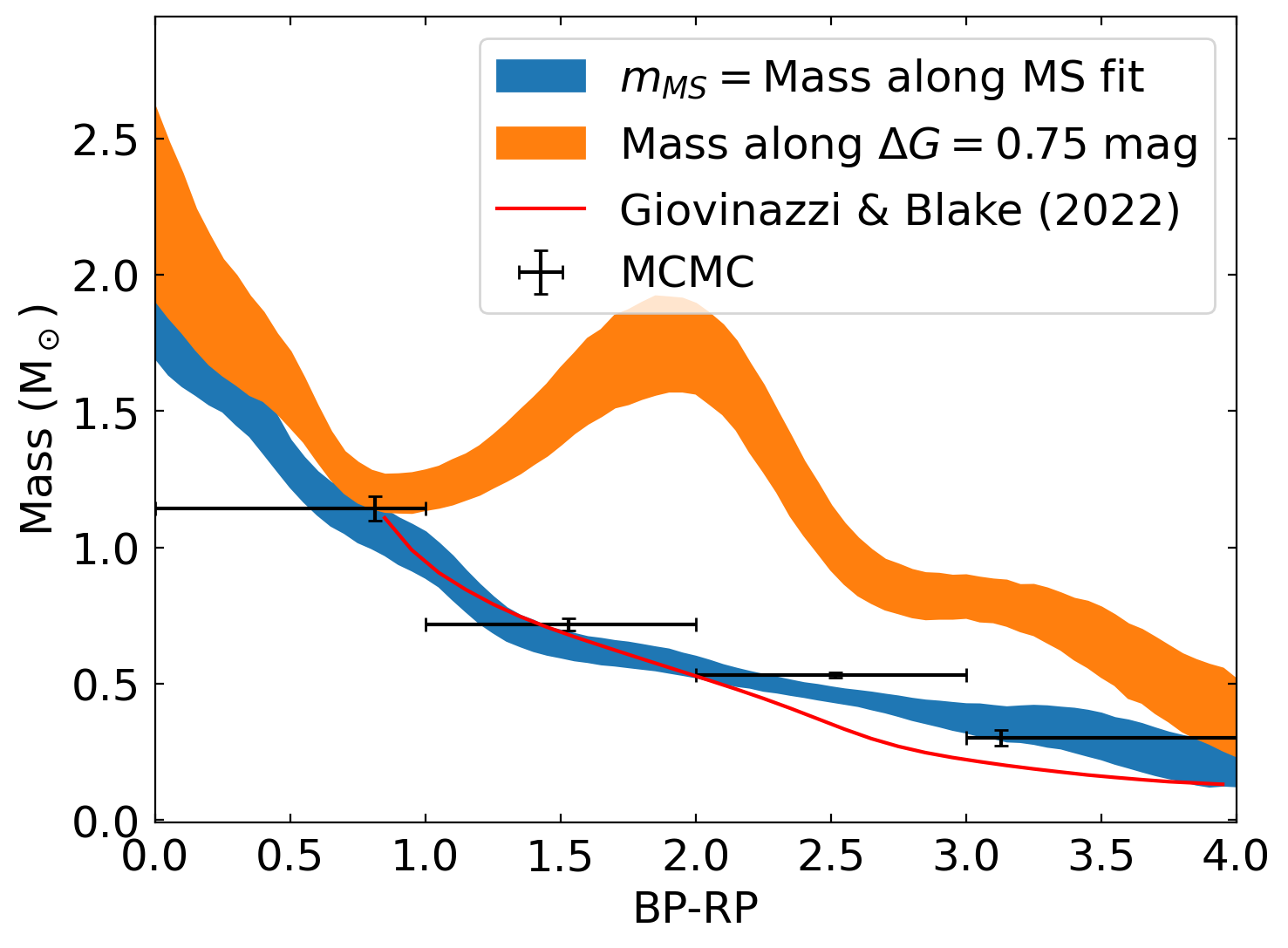}
    \caption{Left: the distribution of $p(\tilde u)$ from the MCMC method. Right: comparison of mass measurements from the neural-network method (blue and orange line, same as in Fig.~\ref{fig:gaia-HR-twinMS}), the MCMC method (black crosses), and the functional fit from \cite{Giovinazzi2022} (red line). }
    \label{fig:mcmc-comparison}
\end{figure*}

Fig.~\ref{fig:mcmc-corner} shows the posterior distributions (made using \texttt{corner}, \citealt{Foreman-Mackey2016}) of the mass parameters. This result consists of 32 walkers, each with 5000 steps. The computation time takes about 1.5 hours on the Google Colab platform (no GPU is used). The 16-50-84 percentiles of masses are:
\begin{align}
    m[0] =& 1.142_{-0.045}^{+0.045}\ {\rm M_\odot} \\
    m[1] =& 0.718_{-0.021}^{+0.022}\ {\rm M_\odot} \\
    m[2] =& 0.532_{-0.011}^{+0.011}\ {\rm M_\odot} \\
    m[3] =& 0.302_{-0.030}^{+0.029}\ {\rm M_\odot} \\
    m[4] =& 0.631_{-0.059}^{+0.063}\ {\rm M_\odot}
\end{align}

These measurements are consistent with the literature. For example, the mean white dwarf mass from MCMC is $m[4] =0.631_{-0.059}^{+0.063}$\Msun, consistent with the mean mass of $0.647^{+0.013}_{-0.014}$\Msun\ DA white dwarfs using gravitational redshifts \citep{Falcon2010}. The left panel of Fig.~\ref{fig:mcmc-comparison} is the resulting distribution of $p(\tilde u)$ where $\tilde u$ is computed using the 50-percentile masses, showing that the result is in good agreement with the model. Also, the high-$\tilde u$ feature is similar to the one from the neural-network method (Fig.~\ref{fig:gaia-HR-pu}).

The right panel of Fig.~\ref{fig:mcmc-comparison} compares the MCMC results with the neural network results from Fig.~\ref{fig:gaia-HR-twinMS}. The horizontal values of the MCMC results (black crosses) are the median BP$-$RP in the sample, and their horizontal bars indicate the range of the BP$-$RP bins. The vertical bars are the 1-sigma mass uncertainties. Compared to the neural network's single-star results (blue line), the MCMC masses are slightly overestimated because the MCMC sample includes both single stars and unresolved binaries. We also overplot the result from \cite{Giovinazzi2022} as the red line, where the authors use twin wide binaries to derive the relation between mass and the absolute RP magnitudes. We apply their fit (their Eq. 8) to derive the masses for the field MS stars, and then the red line in Fig.~\ref{fig:gaia-HR-twinMS} is the median mass binned by their BP$-$RP colors. The blue and red lines overall agree with some noticeable differences. We are cautious that this comparison is mainly for illustration, and the detailed difference is subject to the photometry conversion, the choice of the fitting function used in \cite{Giovinazzi2022}, and the different wide binary samples.

The advantage of MCMC is that it provides robust uncertainties for the masses. Using only 10,000 (8\%) out of 126,000 wide binaries, we can reach mass uncertainties down to $0.01$-$0.07$\Msun. These uncertainties give us a sense of mass uncertainties in the neural-network method, which considers a larger sample but smaller-scale structures (smaller bins in the MCMC sense, although the neural network does not explicitly perform binning). The disadvantage of MCMC is that it is time-consuming. With only 8\% of the sample and only 5 parameters for the model, MCMC already takes 1.5 hours to converge. If we want to explore structures in the mass measurements of the entire H-R diagram, MCMC requires much more ($>50$) mass indices and its convergence becomes challenging within a reasonable computation time. This is the motivation for us to develop a neural network-based method in the main text.

\bibliography{paper-massHR}{}
\bibliographystyle{aasjournal}
\end{document}